\documentclass{aa}  

\usepackage{graphicx}
\usepackage{txfonts}
\usepackage{lipsum}
\usepackage{subcaption}         
\usepackage{lscape}           
\usepackage{placeins}
\usepackage{booktabs}
\usepackage{amsmath}
\usepackage{tabularx}
\usepackage{xcolor}

\usepackage[colorlinks=true,
linkcolor=blue,
citecolor=blue,
urlcolor=blue]{hyperref}

\begin{document}

        \title{Shape of the direct‑method mass-metallicity relation with JWST}
     \subtitle{Fast‑track nitrogen and helium enrichment}

        \author{A.~Giménez-Alcázar\inst{1},
    R.~Amorín\inst{1},
    J.M.~Vilchez\inst{1}
        }
        
        \institute{
                Instituto de Astrofísica de Andalucía (IAA-CSIC), Glorieta de la Astronomía s/n, 18008 Granada, Spain\\ \email{agimenez@iaa.es} }
        
        \date{Accepted September 02, 2026}
        
        \abstract
{}
{We investigate the mass-metallicity relation (MZR) from $z = 1$ to $z = 9$ using electron-temperature based gas-phase metallicities. We explore how auroral-line selection effects, SFHs, and secondary abundances impact its interpretation in the early Universe.}
{We compiled a homogeneous sample of 286 star-forming galaxies observed with JWST/NIRSpec medium-resolution spectroscopy, selected through detections of the [O\,III]\,$\lambda$4363 auroral line from the public DAWN JWST Archive (DJA). We derived the electron densities, temperatures, and oxygen abundances using the direct $T_{\rm{e}}$ method, along with relative N/O and He/H abundances. Stellar masses are obtained via SED fitting and star formation rates from reddening-corrected Balmer emission lines. To quantify any auroral-line selection biases, we additionally stacked sets of galaxy spectra with and without auroral-line detections, extending the MZR into regimes that are inaccessible with respect to individual measurements.}
{The auroral line-detected sample spans stellar masses of $\log(M_\star/M_\odot) = 6.77$--$10.5$, star formation rates of $\log(\mathrm{SFR}\,[M_\odot\,\mathrm{yr}^{-1}]) = -0.24 \text{ to } 2.38$, and low metallicities of $12 + \log(\mathrm{O/H}) = 6.9$--$8.4$, with electron temperatures of T[O\,III] = 10030 K to 37400 K and electron densities ranging from 100$-$990\,cm$^{-3}$. A linear fit to the individually detected sample yields a MZR slope of $\gamma = 0.38 \pm 0.09$. Stacked galaxies without individual auroral-line detections define a relation with a similar slope, but a systematically higher metallicity by ~0.2-0.3 dex at a fixed stellar mass. These two sequences differ markedly in their star-formation properties: galaxies with individual auroral-line detections exhibit higher SFRs, larger emission-line equivalent widths, and larger deviations from the fundamental metallicity relation (FMR). In contrast, stacked non-detections appear more chemically evolved and closer to the FMR. In several stacked bins, we also found enhanced N/O and He/H ratios, suggesting rapid enrichment channels such as Wolf-Rayet stars or supermassive stars.}
{The MZR of high-z galaxies traced by JWST at low stellar masses is shaped both by recent SFH and auroral-line selection effects. Auroral lines preferentially identify high-EW and high-sSFR galaxies in the low-metallicity envelope of the relation, where the effects of dilution due to inflows of pristine gas are greater; whereas non-detections reveal a more enriched sequence closer to the metallicity predicted by the FMR.}

        
        \maketitle
    \nolinenumbers

        \section{Introduction}
        
        Scaling relations among global galaxy properties provide fundamental constraints on the processes that drive galaxy formation and evolution. These empirical correlations reveal underlying physical connections between mass assembly, star formation, and structural growth. One of the most fundamental correlations in galaxy evolution is the relation between stellar mass and gas-phase metallicity, commonly known as the mass–metallicity relation (MZR). Metallicity acts as a fossil record of a galaxy's star formation history (SFH) and gas exchange processes, making it a valuable observable for understanding the evolutionary processes of a galaxy. Since the pioneering work of \citet{Lequeux1979} for star-forming galaxies (SFG), subsequent studies have established the MZR as a cornerstone in galaxy evolution research.\\
        
        At low redshift, the MZR relation has been characterised in detail using large spectroscopic samples. The Sloan Digital Sky Survey (SDSS) provided the first robust statistical description of this relation for SFG in the Local Universe \citep{Tremonti2004}. Subsequent studies have refined its calibration and investigated its secondary dependencies, including the role of star formation rate (SFR), environment, and internal metallicity gradients \citep[e.g.][]{Mannucci2010,Berg2012,Andrews2013,Zahid2014, Curti2020,Duarte2022A&A}.In particular, \citet{Ellison2008ApJ...672L.107E}, \citet{Mannucci2010}, and \citet{Lara-lopez2010A&A...521L..53L} proposed that the stellar mass, gas-phase metallicity, and SFR define a tight three-dimensional relation, known as the fundamental metallicity relation (FMR), whereby galaxies with higher SFRs tend to exhibit lower metallicities at fixed stellar mass. The FMR has been interpreted as evidence that gas accretion and star formation are closely coupled in regulating galaxy chemical evolution. Subsequently, integral-field spectroscopy programmes such as CALIFA and MaNGA have expanded this picture by resolving metallicity gradients and connecting local chemical properties to global scaling relations. These observations offer hints about the physical processes that regulate the MZR, including gas inflows, outflows, and feedback that shape the MZR \citep[e.g.][]{Sanchez2017,Belfiore2017,Maiolino2019A&ARv..27....3M,Kewley2019ARA&A..57..511K}.\\

        Beyond the Local Universe, numerous studies have investigated the evolution of the MZR with cosmic time. Galaxies at higher redshift might exhibit lower metallicities at fixed stellar mass, a trend that is often interpreted as reflecting the progressive chemical enrichment of the interstellar medium (ISM). This apparent evolution (0.2-0.3 dex) has been reported up to $z \sim 3$ using rest-frame optical spectroscopy from large surveys such as KBSS, MOSDEF, and FMOS-COSMOS, providing constraints on the MZR and its possible dependence on SFR and ionisation conditions \citep[e.g.][]{Erb2006,Maiolino2008A&A...488..463M,Zahid2013,enrique2013,Steidel2014,Troncoso2014,Sander2015,Onodera2016,Strom2017, Curti2020,Sanders2021,Strom2022,Korhonen2025}. However, these early efforts faced significant limitations that complicate the interpretation and evolution of the MZR. High‑redshift spectroscopic samples were typically biased towards relatively massive galaxies, leaving the low‑mass regime ($M_\star \lesssim 10^{9}M_\odot$) poorly constrained even in the Local Universe \citep[e.g.][]{Henry2013}. Moreover, the heterogeneous use of strong‑line (SL) abundance diagnostics introduced systematic offsets comparable to the reported cosmic evolution of the relation \citep{Kewley2008,Ly2016ApJ...828...67L,Calabro2017A&A...601A..95C}. Most calibrations were anchored to local galaxy populations and it is only recently that new calibrations have begun to incorporate systems with auroral-line detections at high redshift. These temperature-sensitive lines make it possible to estimate the electron temperature of the ionised gas, from which ionic abundances can be derived through the so-called direct (or Te) method. In this approach, the temperature is determined from ratios involving an auroral line and the corresponding stronger nebular lines, providing a physical estimate of the gas-phase oxygen abundance \citep[e.g.][]{Perez-montero2017PASP..129d3001P,Maiolino2019A&ARv..27....3M}. In addition, existing SL calibrations are far from uniform: they rely on non-homogeneous datasets with differing selection functions and employ a variety of line‑ratio combinations. Many commonly used diagnostics are sensitive not only to O/H but also to secondary parameters such as the ionisation parameter ($U$), electron temperature, and the nitrogen‑to‑oxygen ratio (N/O; \citealt{montero2014MNRAS.441.2663P,Maiolino2019A&ARv..27....3M,enrique2021MNRAS.504.1237P}). As a result, systematic biases, especially in the low-mass, low-metallicity regime, are unavoidable. Several indicators (e.g. R23, R2, R3)\footnote{Here, R23 denotes ([O\,II] $\lambda\lambda3726,3729$ + [O\,III] $\lambda\lambda4959,5007$)/H$\beta$, while R2 and R3 refer to [O\,II] $\lambda\lambda3726,3729$/H$\beta$ and [O\,III] $\lambda\lambda4959,5007$/H$\beta$, respectively.} also become degenerate at intermediate metallicities, requiring additional constraints to break these degeneracies \citep[e.g.][]{Perez-montero2005,Moustakas2010}. Strong‑line techniques could approach the accuracy of the Te method only when multiple line ratios are available.\\

    The earliest phases of galaxy evolution remained largely unexplored until the advent of JWST. Its sensitivity now enables metallicity measurements for galaxies at $z > 3$ and, for the first time, the exploration of the MZR deep into the epoch of reionisation. At high redshift, the faintness of the [O\textsc{iii}]$\lambda4363$ auroral line restricts the use of the direct electron‑temperature ($T_{\rm{e}}$) method. As a result, large galaxy samples still rely on SL diagnostics based on optical nebular emission lines \citep[e.g.][]{Matthee2023,Nakajima2023ApJS..269...33N,Li2023,Curti2024,Chemerynska2024,roland2025,Stanton2025arXiv251100705S}. However, the Te direct method, is far less model dependent and is therefore more reliable. Given its importance, an increasing number of dedicated JWST spectroscopic programs such as AURORA, MARTA, and CECILIA have been designed specifically to probe the detectability of the [O\textsc{iii}]$\lambda$4363 auroral line, which provides more robust abundance estimates for galaxies up to $z \sim 9$ \citep[e.g.][]{Strom2023,Sanders2025arXiv250810099S,cataldi2025}.\\
        
        Despite the progress made in recent years, several important questions remain unclear. For example, an apparent redshift evolution in the shape of the MZR has been suggested for relatively massive galaxies ($M_\star > 10^{9}\,M_\odot$). However, it remains unclear whether low-mass galaxies ($M_\star < 10^{9}\,M_\odot$) follow a similar evolutionary trend. Some studies suggest that the relation flattens at high redshift \citep[e.g.][]{Li2023,Curti2024}, while others find steeper trends \citep[e.g]{Chemerynska2024,Raptis2025}. Simulation-based MZR predictions are similarly unsettled, showing their largest discrepancies at the low-mass end across redshift \citep[e.g.][]{Marszewski2024ApJ...967L..41M,Garcia2025MNRAS.536..119G,Sharda2026}. In addition, sample selection effects driven by the detection limits of auroral lines, required to probe the MZR at low metallicities, may introduce biases in such evolutionary studies. Interpreting the MZR in the early Universe is therefore challenging, resulting in a wide range of MZR shapes in current JWST samples at high redshift $z\gtrsim1$–10. A complementary perspective comes from the gas-regulator models, which provide a natural physical framework for understanding the connection between stellar mass, metallicity, and star-formation rate. In these models, the gas reservoir of a galaxy is set by the balance between gas accretion, star formation, and outflows \citep[e.g.][]{Lilly2013} following the FMR \citep[e.g.][]{Mannucci2010,Lara-lopez2010A&A...521L..53L,Andrews2013,Curti2020}. Deviations from the FMR should reflect departures from equilibrium, driven by changes in the processes regulating the gas cycle. Observational work indicates that the FMR holds no evolution up to $z\sim2.5$ for relatively massive galaxies ($M_\star > 10^{9}\,M_\odot$). However, its validity for low-mass systems remains debated \citep[e.g.][]{cresci2012MNRAS.421..262C,yates2012MNRAS.422..215Y,Curti2020,topping2021MNRAS.506.1237T}. Extending the comparison to the early Universe has yielded mixed results. Several studies report that galaxies at $z\gtrsim4$ appear significantly more metal-poor than predicted by the $z=0$ FMR \citep[e.g.][]{Nakajima2023ApJS..269...33N,Curti2024,Heintz2024Sci...384..890H,Pollock2025,Stanton2025arXiv251100705S,Sarkar2025ApJ...978..136S,kotiwale2026A&A...706A.165K}, while others find little to no offset \citep[e.g.][]{Faisst2025,roland2025}\\
        
        In this work, we provide new insights into the shape of the MZR and explore how it relates to other global galaxy properties using a large sample of sources at $1<z<9$ drawn from the DAWN JWST Archive (DJA)\footnote{https://dawn-cph.github.io/dja}. These galaxies are identified by detecting the [O\,III]$\lambda4363$ auroral line, which enables direct metallicity measurements. Unlike comparisons based on heterogeneous literature samples, all galaxies in our analysis were processed and measured in a homogeneous manner using the same pipeline. This allowed us to reduce any systematic uncertainties arising from differences in data reduction, line-flux measurements, and abundance determinations. This approach has offered us a reliable view of the low‑mass end of the MZR and enabled an investigation of how its scatter depends on SFR. In parallel, we also analysed galaxies without individual [O\,III]$\lambda$4363 detections by constructing stacked spectra, extending the MZR to regions of parameter space where auroral lines are not measurable via a galaxy-by-galaxy approach. The comparison between detections and stacks enabled us to explore whether any differences can be interpreted within the framework of equilibrium gas‑regulator models and to assess how they relate to the position with respect to the FMR. Finally, we examined nitrogen and helium abundances as complementary tracers of chemical enrichment beyond O/H. Overall, Nitrogen and helium are produced through partly distinct channels and on different timescales, so together they discriminate among enrichment scenarios that N/O alone cannot resolve. Enrichment by Wolf-Rayet (WR) stars or very massive stars \citep[e.g.][]{Kobayashi2020ApJ...900..179K} is expected to raise both N/O and He/H simultaneously, whereas delayed secondary production by intermediate-mass stars raises N/O at essentially constant He/H \citep[e.g.][]{Berg2019,Nichols10.1093/mnras/stw3235}. Dilution by accretion of pristine gas raises N/O only by depressing O/H. Helium is particularly informative in this regard because its enrichment is measured as a small excess above the primordial value and is therefore sensitive to recent massive-star nucleosynthesis rather than to the integrated SFH. \\
    
        The paper is organised as follows. Section 2 describes the sample selection and observations. Section 3 explains the data reduction and estimation of physical parameters. Section 4 presents our results. Section 5 discusses the shape of the MZR and the role of the SFR. Section 6 presents the nitrogen and helium enrichment of our sample, while Sect. 7 summarises our conclusions. In the following, we adopt the $\Lambda$CDM cosmological model with parameters $H_0 = 70\,\mathrm{km\,s^{-1}\,Mpc^{-1}}$, $\Omega_{M} = 0.3$, and $\Omega_{\Lambda} = 0.7$. We adopt the solar abundance scale of \citet{Asplund2021A&A...653A.141A}, with  $12+\log({\rm O/H})_\odot = 8.69$ and $12+\log({\rm N/H})_\odot = 7.83$. This corresponds to a solar abundance ratio of $\log({\rm N/O})_\odot = -0.86$.

        \section{Data selection}

We used JWST NIRSpec micro-shutter assembly (MSA) spectroscopy compiled from DJA. We selected the latest public NIRSpec datasets (v4), focusing only on sources observed with medium–high spectral resolution and excluding prism spectroscopy. We also imposed a so-called grade > 2 criterion to ensure robust redshifts derived from one or more emission or absorption features. Finally, we used the DJA catalogue to select only galaxies with a detection of [O\,III]$\lambda$4363 with S/N > 2, resulting in a sample of 1546 galaxies between redshift z$\sim$1$-$9. We adopt S/N $\geq 2$ as a compromise between sample size and reliability; more than 60\% of the sample has S/N $> 3$ in [O\, III]$\lambda 4363$. All $\lambda 4363$ detections were visually inspected to verify that the emission is genuine and not driven by artefacts, and their uncertainties are propagated throughout the analysis. We explain in Sect.~\ref{mzr} that adopting a stricter S/N $> 3$ selection did not impact our results.

    For further details on the data reduction processes of the spectrum, we refer to  \cite{Graaff2025} and \cite{Heintz2024Sci...384..890H}. For this sample, we also compiled photometric measurements from the DJA Mosaic v7 release, which combines HST and JWST imaging. In this work, we adopted aperture-corrected fluxes matched to the auto flux in the detection band and applied corrections for flux outside the auto aperture for the HST bands. Further details on photometric processing can be found in \citet{Valentino2023ApJ...947...20V}.\\

\begin{figure}
            \centering
            \includegraphics[width=1\linewidth]{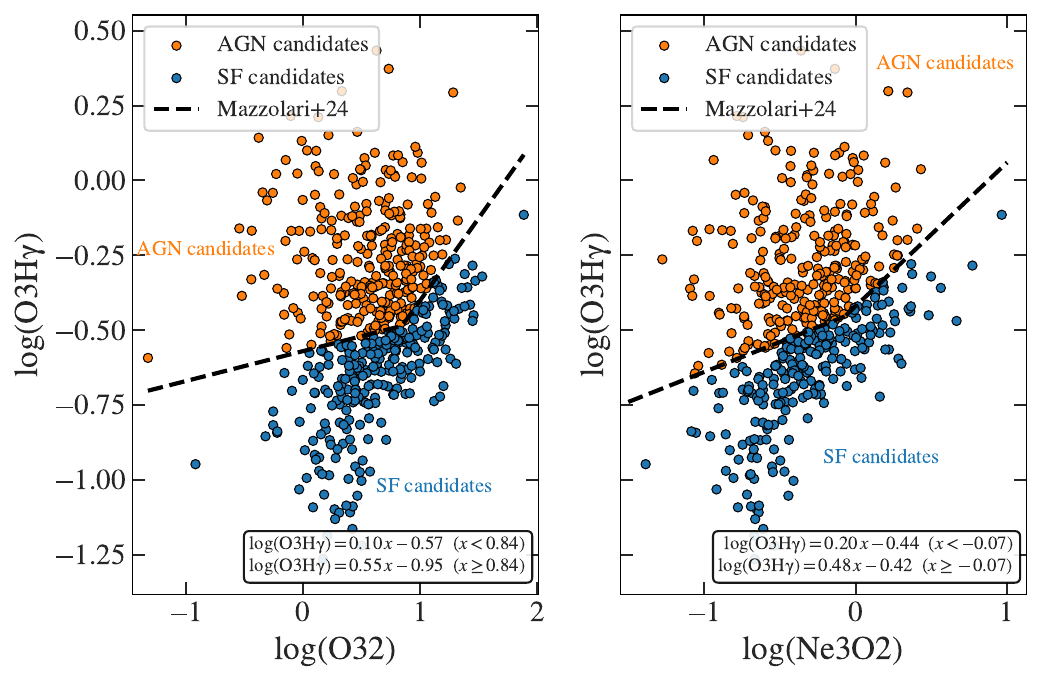}
            \caption{Diagnostic diagrams used to separate AGNs from star-forming candidates.
In all panels, orange and blue symbols indicate AGN and SF candidates, respectively.
Dashed lines show the empirical demarcation criteria of \citet{Mazzolari2024A&A...691A.345M}. See the text for details.}
            \label{fig:mazzolari}
        \end{figure}

    We employed the diagnostic diagrams suggested in \cite{Mazzolari2024A&A...691A.345M} based on the [\ion{O\,III}]$\lambda$4363 auroral line (see also \citealp{Backhaus2025}). Specifically, we employed the diagnostic diagrams based on the auroral ratio, $\mathrm{O3H\gamma} \equiv [\ion{O\,III}]\,\lambda4363 / \mathrm{H}\gamma$, combined with either the excitation-sensitive ratio, $\mathrm{O32} \equiv [\ion{O\,III}]\,\lambda5007 / [\ion{O\,II}]\,\lambda3727$, or the hardness-sensitive ratio, $\mathrm{Ne3O2} \equiv [\ion{NeIII}]\,\lambda3869 / [\ion{O\,II}]\,\lambda3727$. All diagnostics were performed after applying the interstellar extinction correction described in the next section. These demarcations were defined by \citet{Mazzolari2024A&A...691A.345M} based on empirical calibrations using local and high-redshift sources, supported by a broad range of photoionisation models. In Fig.~\ref{fig:mazzolari}, we show how the sample of galaxies is distributed across the two diagnostic diagrams. The boundaries separating these populations are indicated in each plot. We classify each object according to its position in these diagnostic planes and exclude those consistent with active galactic nucleus (AGN) ionisation. For all the diagnostic diagrams, we required a signal-to-noise ratio greater than 2 in all the emission lines involved.\\

\begin{figure}
                \centering
                \includegraphics[width=1\linewidth]{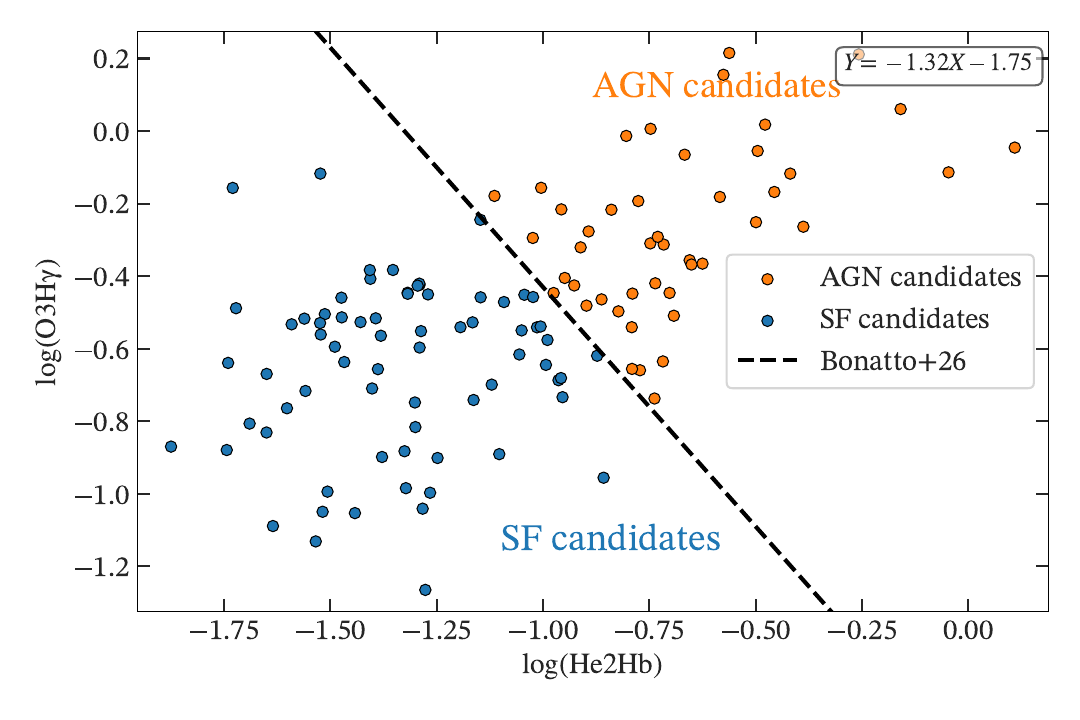}
                \caption{Diagnostic diagram based on the emission-line ratios $\log(\mathrm{He2Hb})$ and $\log(\mathrm{O3H\gamma})$. Orange points represent AGN candidates, while blue points correspond to star-forming galaxies. The dashed~black  line shows the empirical demarcation defined by \citet{bonatto2026} based on local (z<1) analogues with DESI.}
                \label{fig:He2Hb_O3Hg_diagram}
        \end{figure}

        Another powerful diagnostic is accessible through the \ion{He\,II}\, emission. \ion{He\,II}\, is particularly sensitive to the hardness of the ionising spectrum, making it an excellent tracer of non-stellar excitation such as AGN activity or fast radiative shocks \cite[e.g][]{Brinchmann2008,Shirazi2012}. We used the diagnostic ratio $\mathrm{O3H\gamma/He2Hb}$ defined in \citet{bonatto2026}, where $\mathrm{He2Hb} \equiv [\ion{He\,II}]\,\lambda4686 / \mathrm{H}\beta$. In Fig. \ref{fig:He2Hb_O3Hg_diagram} We show all the objects with He\,II emission and their positions in the diagnostic diagram, together with the adopted selection criterion. Finally, we removed any detection (even if marginal) of \ion{NeV} \,$\lambda$3426, typical of narrow-line AGNs from the sample presented in \citep[e.g.][]{Scholtz2025A&A...697A.175S}. For this study, we adopted the union of the star‑forming regions across the three diagnostic diagrams: a source was retained if it was classified as star-forming (SF) in at least one diagram, rather than applying the stricter intersection of all three. After that, we used the mass–excitation (MEx) diagram \citep{Juneau2014ApJ...788...88J}, relating stellar mass and emission-line excitation, with the high-redshift modification of \citep{coil2015ApJ...801...35C}, to check whether our sources were lying in the SF region. Although the diagnostics do not absolutely guarantee the absence of AGNs in the sample, we found no clear evidence of their presence.\\
    
    The final sample of galaxies with detected [\ion{O\,III}] auroral emission consists of 286 objects spanning a redshift range of 0.8 to 9.1. The programmes and parent catalogues contributing to this sample are listed in Table~\ref{tab:survey_counts}. 
        
        \begin{table}[htbp]
\centering
\caption{Number of objects per survey and corresponding reference.}
\label{tab:survey_counts}
{\footnotesize
\renewcommand{\arraystretch}{0.9}
\begin{tabular}{l c l}
\hline
Survey & N Objects & Reference \\
\hline
AURORA & 18 & \cite{Shapley2025ApJ} \\
Bluejay & 13 & \cite{Belli2024Natur} \\
CECILIA & 1 & \cite{Strom2023} \\
CEERS-ERS & 11 & \cite{Finkelstein2023} \\
DEEPDIVE & 7 & \cite{2025arXiv250622642I} \\
EXCELS & 30 & \cite{Carnall2024} \\
GLASS-ERS & 2 & \cite{Mascia2024} \\
GLIMPSE & 5 & DD 9223 \\
P. Arrabal Haro  & 1 & GO 2674 \\
J.~Antwi-Danso  & 1 & GO 4318  \\
GTO (C. Willott) & 2 & \cite{Christensen2023} \\
GTO (G. Rieke) & 8 & GTO 1207 \\
GTO macs1149 & 14 & \cite{Stiavelli2023} \\
GTO Wide & 1 & GTO 1215 \\
JADES & 111 &  \cite{Eugenio2025}\\
JADES Ultra-Deep & 5 & \cite{Eisenstein25ApJ} \\
LyC-22 & 11 & GO 1869 \\
MARTA & 16 & \cite{cataldi2025} \\
RUBIES & 8 & \cite{Graaff2025} \\
S.~Fujimoto & 9 & GO 4762 \\
SMACS-0723 ERO & 3 & \cite{Pontoppidan2022ApJ} \\
STARK & 4 & \cite{Topping2025MNRAS.541.1707T} \\
SUSPENSE & 1 & \cite{Slob2024ApJ} \\
- & 3 & \cite{Wang2024} \\
- & 1 & \cite{Williams2023} \\
- & 2 & \cite{Chisholm2024MNRAS} \\
- & 2 & \cite{Frye2024ApJ} \\
- & 2 & \cite{Maseda2023} \\
- & 2 & \cite{Nakajima2025} \\
- & 2 & \cite{Tang2025} \\
\hline
\end{tabular}
}
\end{table}

        \section{Estimation of physical properties of the sample}
        Having defined a sample of star-forming galaxies, we now turn to the derivation of their physical properties. We focus on key parameters such as extinction, density, electron temperature, and chemical abundance, derived from the measured emission line fluxes. At the end of this section, we will also derive photometric properties through spectral energy distribution (SED) fitting.
        
        \subsection{Extinction}
        \label{sec:dust}
        
        The fluxes of emission lines can be affected by dust extinction and correcting for this effect is essential to accurately estimate the gas-phase metallicity of galaxies. We calculate the de-reddened fluxes assuming the \cite{Cardelli1989ApJ...345..245C} nebular attenuation curve with $R_V = 3.1$. We follow the procedure developed in \citet{bonatto2026}.
        
        The nebular extinction coefficient, $c(\mathrm{H}\beta)$, for each galaxy was computed using multiple Balmer line ratios: H$\alpha$/H$\beta$, H$\gamma$/H$\beta$, and H$\delta$/H$\beta$. For each ratio, we performed a Monte Carlo simulation to propagate the measurement errors. Ratios where either line had a S/N below a given threshold (i.e. $\mathrm{S/N} < 2$) were discarded. From the remaining valid ratios, we selected a final $c(\mathrm{H}\beta)$ with the smallest uncertainty estimated from the Monte Carlo distribution. The theoretical intrinsic ratios assume Case~B recombination at $T_{\rm{e}} = 15{,}000$\,K and $n_{\rm{e}} = 100$\,cm$^{-3}$. 
    
    For a given Balmer ratio $(\lambda_1/\lambda_2)$, the extinction coefficient can be calculated as
        \begin{equation}
                c(\mathrm{H}\beta)_{(\lambda_1/\lambda_2)} = \frac{\log\left[\dfrac{(F_{\lambda_1}/F_{\lambda_2})_{\text{obs}}}{(F_{\lambda_1}/F_{\lambda_2})_{\text{int}}}\right]}{f(\lambda_1) - f(\lambda_2)},
        \end{equation}
        where $(F_{\lambda_1}/F_{\lambda_2})_{\text{obs}}$ is the observed flux ratio, $(F_{\lambda_1}/F_{\lambda_2})_{\text{int}}$ is the theoretical intrinsic ratio for Case~B recombination, and $f(\lambda)$ is the reddening curve normalized at H$\beta$ (i.e. $f(\mathrm{H}\beta) = 0$).  For each galaxy, a $c(\mathrm{H}\beta)$ is adopted as the value that minimises the uncertainty of the MC distribution. When no positive values are obtained in some cases, we set the extinction to zero, $c(\mathrm{H}\beta)$=0. Although part of the discrepancy may be due to measurement uncertainties, such low values can arise from differences in the actual T$_e$ and n$_e$ compared to the assumed theoretical conditions, or from deviations from pure case‑B recombination or Balmer self‑absorption. We find a median extinction coefficient $c(\mathrm{H}\beta)_{med}$= 0.21 $\pm 0.09$. The observed flux $F_\lambda$ is then corrected using        \begin{equation}
                I_\lambda = F_\lambda \times 10^{c(\mathrm{H}\beta)\, f(\lambda)}
        ,\end{equation}
        where $I_\lambda$ is the intrinsic flux. We adopt the \citet{Cardelli1989ApJ...345..245C} extinction law with RV=3.1 for $f(\lambda)$. Across the wavelength range of our data (optical rest‑frame), the specific choice of attenuation curve only has a minor effect on the resulting line ratios. Using the \citet{calcetti2000ApJ...533..682C} starburst law or the high‑z prescription from \citet{Reddy2020}, for instance, leads to variations that remain well within our observational uncertainties. We further assume a unity dust covering fraction ($f_{\rm cov}=1$). As recently shown by \citet{Reddy2026ApJ...999...15R}, $f_{\rm cov}<1$ could lead to either an underestimation of the nebular extinction. However, breaking this degeneracy requires joint Balmer and Paschen recombination line measurements, which are not available for our sample.
        
        \subsection{Electron density}
        
        The [O\,II] $\lambda\lambda3727,3730$ doublet, used as a density diagnostic for the low-ionisation zone, cannot be resolved with medium-resolution NIRSpec gratings. Because our data do not contain other density diagnostics for the high-ionisation zone (e.g. [Ar\,IV] or C\,III]), we adopt electron density from [S\,II] $\lambda\lambda6717,6731$ doublet as the representative electron density when deriving $T_{\rm{e}}$. Electron densities are computed with the PyNeb package (v1.1.27; \citealp{Luridiana2015A&A...573A..42L}) from the observed flux ratio: R$_{SII}$ = I(6717)/I(6731). The adopted limits follow the behaviour of PyNeb emissivity curves at $T_{\rm{e}} = 15{,}000$\,K: at both low and high densities, $R_{\text{S\,II}}$ becomes insensitive to $n_{\rm{e}}$. When the measured ratio falls in these flat regions, we assign representative values:
        \begin{itemize}
                \item $R_{\text{S\,II}} \ge 1.389$: low-density plateau, $n_{\rm{e}} \simeq 100~\text{cm}^{-3}$ \citep{Mendezdelgado2023MNRAS.523.2952M}.
                \item $R_{\text{S\,II}} \le 0.951$: high-density regime, $n_{\rm{e}} = 1000~\text{cm}^{-3}$.
        \end{itemize}
        
        Exact densities can be computed with \texttt{PyNeb} only when $0.951 < R_{\text{S\,II}} < 1.389$. If the [S\,II] lines are undetected with ${\rm S/N} < 2$ or unavailable, we adopted a default density of $n_{\rm{e}} = 300~\text{cm}^{-3}$, a value commonly used for high-redshift galaxies \citep[e.g.][]{Sanders_2016,Matthee_2023,Curti2024,Sanders2025arXiv250810099S,Topping2025MNRAS.541.1707T}. The median $n_{\rm{e}}$ for galaxies with measured $R_{\text{S\,II}}$ is 390~cm$^{-3}$, consistent with these studies and expanding from 102 to 990 cm$^{-3}$.

        \subsection{Electron temperature}
        
        We assumed a two-zone nebula with two corresponding electron temperatures for the high and low ionisation zones. The electron temperature of the high-excitation regions, T[{O\,III}], was computed from the flux ratio $(\lambda4959 + \lambda5007)/\lambda4363$ using \texttt{PyNeb}. Monte Carlo simulations were performed to propagate the observational uncertainties in the line fluxes. For low-excitation regions, the electron temperature T[{O\,II}] was then estimated from T[{O\,III}] using the density-dependent relation of \cite{Hagele2006MNRAS.372..293H}; noting that this prescription was applied only when a reliable density estimate was available (i.e.\ $n_{\rm{e}}\approx 100$--$1000\ \text{cm}^{-3}$, provided by the [S\,II] doublet).

        \begin{align}
                T[{O\,II}] = \dfrac{1.2 + 0.002\,n_{\rm{e}} + 4.2/n_{\rm{e}}}{(1/T_h) + 0.08 + 0.003\,n_{\rm{e}} + 2.5/n_{\rm{e}}} \times 10^4\ {K}
        \end{align}
        
        where $T_h \equiv T[{O\,III}]$ is the temperature derived from the [O\,III] auroral-to-nebular line ratio. In the remaining cases, we followed the relation of \cite{stasinka1990A&AS...83..501S} to compute the low-ionisation temperature,
        \begin{align}
                T_{\rm{e}}(\text{O\,II}) = \frac{2.0}{(1/T_h) + 0.8} \times 10^4\ \text{K}.
        \end{align}

    \noindent The distribution of electron temperatures derived from T[{O\,III}] spans a wide range, with the $5^{\rm{th}}$ and $95^{\rm{th}}$ percentiles at $\sim 12,600$K and $\sim 28,300$ K, respectively, and a median value of $\sim 17,100$ K, with a few galaxies above 30,000 K.
        
        \subsection{O/H Abundance}
    \label{O/H method}
        Once the electron temperatures T[{O\,III}] and T[{O\,III}], along with the electron density $n_{\rm{e}}$, have been determined, we can compute the ionic abundances of oxygen directly from the measured line intensities. For $\text{O}^{++}$, we used the combined flux of [O\,III] $\lambda4959$ and $\lambda5007$, together with T[O\,III], following the empirical relation from \cite{Perez-montero2017PASP..129d3001P} and adopting the default collision strengths from \cite{Pradhan2006MNRAS.366L...6P} and \cite{tayal2010ApJS..188...32T}. This is expressed as
        
        \begin{align}
                12 + \log \left( \frac{O^{++}}{H^+} \right) &= 
                \log \left( \frac{I_{4959} + I_{5007}}{I_{H\beta}} \right) 
                + 6.1868 \notag \\
                &\quad + \frac{1.2491}{t_h} 
                - 0.5816 \cdot \log(t_h).
        \end{align}

        Similarly, for O$^+$, the [O II] $\lambda\lambda$3726,3729 doublet and T[{O\,II}] were used, expressed as
        \begin{align}
                12 + \log \left( \frac{O^{+}}{H^+} \right) &= 
                \log \left( \frac{I_{3726}+I_{3729}}{I_{H\beta}}\right) + 5.887 \notag\\
                &\quad + \frac{1.641}{t_l} - 0.543 \log_{10} t_l + 0.000114\,n_{\rm{e}},
        \end{align}
        where $t_l$ = T[{O\,II}] / 10$^4$\ $\mathrm{K}$. Finally, assuming that higher ionisation states (e.g. O$^{3+}$) are negligible, the total oxygen abundance is
        \begin{equation}
                \frac{\mathrm{O}}{\mathrm{H}} = \frac{\mathrm{O}^+}{\mathrm{H}^+} + \frac{\mathrm{O}^{++}}{\mathrm{H}^+}.
        \end{equation}

    The 5th and 95th percentiles of the distribution are $12+\log(\mathrm{O/H}) \simeq 7.26$ and $8.21$, respectively, although a few galaxies exhibit metallicities below the 5th‑percentile limit (e.g. down to $12+\log(\mathrm{O/H}) \approx 6.9$). The median abundance is $12+\log(\mathrm{O/H}) = 7.86$, corresponding to $\sim 15\%$ of the solar metallicity.
        
        \subsection{SED fitting}
    \label{sed_fitting}
        To compute the physical properties, we used the Code Investigating GALaxy Emission \citep[\texttt{CIGALE v2025.0};][]{cigale2019A&A...622A.103B}. \texttt{CIGALE} models parametric SFHs and estimates physical parameters and uncertainties through a Bayesian approach, constructing probability distribution functions for each parameter by evaluating the $\chi^2$ over the full set of models. We adopted a double-exponential SFH, which allowed us to reproduce both a young stellar population and a potential underlying older component. For the stellar population synthesis, we used the Charlot \& Bruzual (2019) models with the \cite{chabrier2003PASP..115..763C} initial mass function (IMF). Nebular emission was incorporated using CLOUDY models \citep{Ferland2013RMxAA..49..137F}, assuming $f_\mathrm{esc}=0$ (all Lyman continuum photons were reprocessed into Balmer lines) and an electron density of $n_{\rm{e}}=100~\text{cm}^{-3}$. Attenuation was accounted for using the modified \cite{calcetti2000ApJ...533..682C} law. Table~\ref{cigale_tab} summarises the adopted configuration.\\
        
        The photometric dataset  Mosaic v7 from DJA includes imaging from both the JWST and the Hubble Space Telescope (HST). We used the aperture-corrected fluxes given by DJA. The list of bands used can be found in \ref{tab:mosaic_v7_photometry}. For our sample of 286 galaxies, we identified photometric counterparts for 226 objects, for which we performed SED fittings. In addition, stellar masses are available in the literature for 16 further galaxies, increasing the total number of sources with stellar mass estimates to 242.

        \begin{table}[h]
                
                \tiny
                \centering
                \caption{\texttt{CIGALE} parameters.}
                \label{cigale_tab}
                \begin{tabular}{cc}
                        \toprule
                        Stellar Parameters &  \\ \midrule
                        $\tau_{\mathrm{main}}$ {[}Myr{]} & 5, 10, 20, 50 \\ \rule{0pt}{2.3ex}
                        $\tau_{\mathrm{burst}}$ {[}Myr{]}      &  0.5, 0.8, 1, 3 \\ \rule{0pt}{2.3ex}
                        Age main {[}Myr{]} & 30, 50, 100, 200, 500        \\ \rule{0pt}{2.3ex}
                        burst\_age {[}Myr{]} & 1, 3, 6, 10, 12 \\ \rule{0pt}{2.3ex}
                        f\_burst& 0.1, 0.15, 0.25, 0.3, 0.5 \\
                        \midrule
                        Charlot \& Bruzual (2019) &  \\ \midrule 
                        IMF & Chabrier \\ \rule{0pt}{2.3ex}
                        metallicity & 0.0001, 0.001, 0.004, 0.008\\ \rule{0pt}{2.3ex}
                        Upper IMF limit [M$_\odot$] & 100\\

                        \midrule 
                        Nebular parameters &  \\ \midrule 
                        z\_gas & 0.0001, 0.001, 0.004, 0.008   \\ \rule{0pt}{2.3ex}
                        $\log{U}$  & -3.5, -3.0, -2.5, -2.0, -1.5   \\ \rule{0pt}{2.3ex}
                        f$_{esc}$, f$_{dust}$ & 0\\
                        
                        \midrule
                        Extinction parameters &  \\ \midrule
                        $E(B-V)_\mathrm{young}$  & 0.1, 0.2, 0.3, 0.4  \\ \rule{0pt}{2.3ex}
                        $E(B-V)_\mathrm{old\_factor}$    & 0.44, 1     \\ 
                        \noalign{\smallskip}
                        
                        \bottomrule
                \end{tabular}
        \end{table}

        \section{Results}
        
        \subsection{Stellar masses and star-formation rates}
\begin{figure}
                \centering
                \includegraphics[width=1\linewidth]{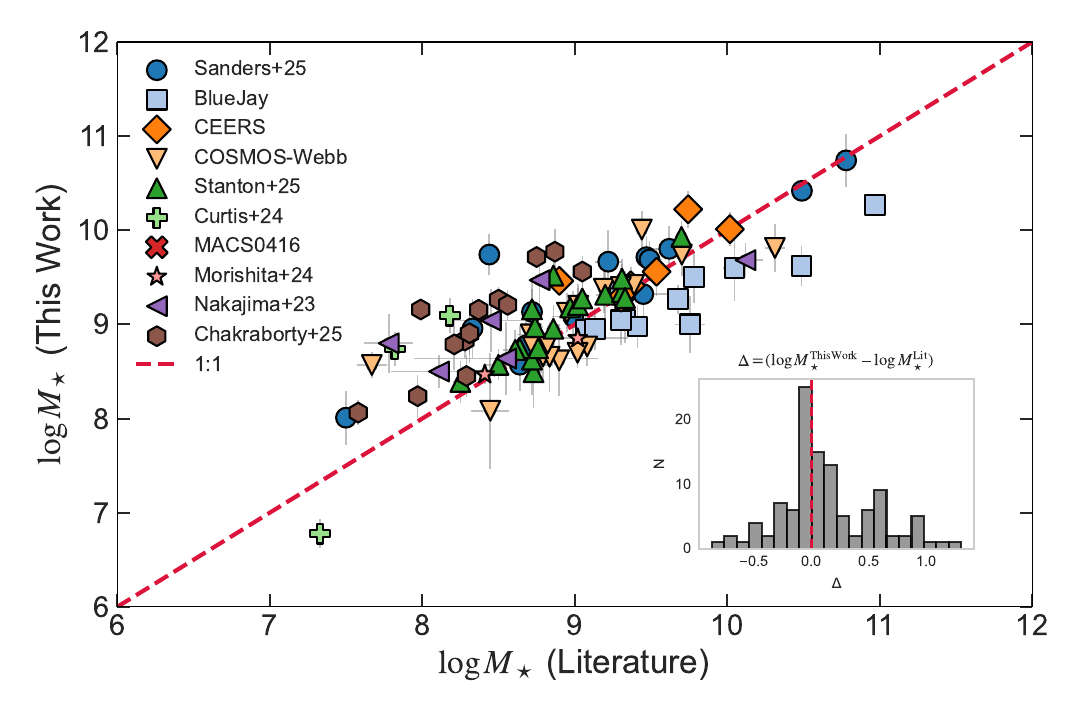}
                \caption{Comparison among stellar mass estimates from this work and values reported in the literature. Each symbol corresponds to a different survey or study, as indicated in the legend. The dashed red line marks the one-to-one relation. Error bars represent uncertainties in both axes. The inset panel shows the distribution of differences, $\Delta = (\log M_\star^{\rm This~Work} - \log M_\star^{\rm Lit})$.}
                \label{fig:mass_comparison_with_literature}
        \end{figure}

    Our sample ranges in stellar masses from  $\log(M_\star/M_\odot)$ =6.77 to 10.74, with the 5th and 95th percentiles at \(\log(M_\star/M_\odot)=8.1\) and 9.8, respectively.  Below $\log(M_\star/M_\odot) <8$, the number of galaxies with auroral line detections drops significantly. Figure~\ref{fig:mass_comparison_with_literature} compares our stellar mass estimates with values reported in the literature. Despite differences in methodology among the studies shown, the discrepancies are generally small, with a standard deviation of only 0.10 dex. However, in some studies, such as \citet{Chakraborty2025}, some individual galaxies show differences of up to 1 dex. These larger discrepancies can be attributed to variations in the assumed SED parameters or the photometry they used. For example, they adopt a constant SFH instead of a double exponential as we do, stellar population synthesis models based on the 2016 update of \citet{bruzal2003MNRAS.344.1000B}, and an initial mass function from \citet{Kroupa2002Sci...295...82K}.\\
        
\begin{figure}
                \centering
                \includegraphics[width=1\linewidth]{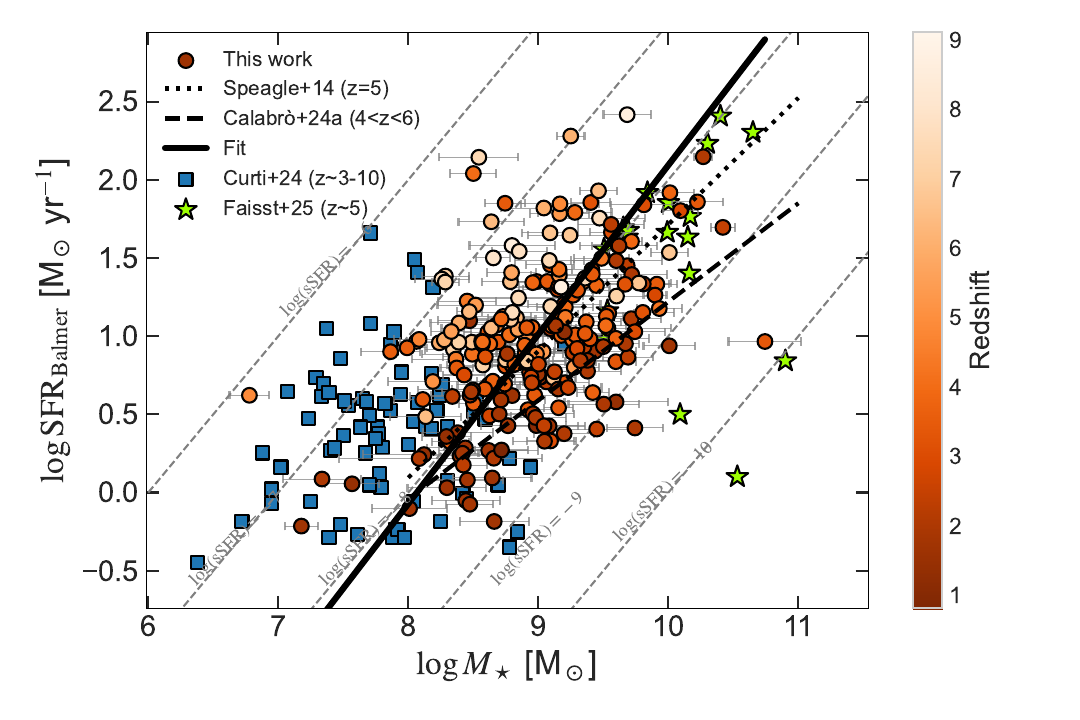}
                \caption{
                        Relation between stellar mass ($\log M_\star/M_\odot$) and SFR ($\log \mathrm{SFR /M_\odot\,\mathrm{yr}^{-1}})$. 
                        Orange points with a colourbar represent the measurements obtained in this work, including error bars in both directions. 
                        Comparisons with main sequence relations reported in the literature are shown: \citet[][$z=5$; dotted black]{speagle2014}, 
                        \citet[][$4<z<6$; dashed black]{Calabro2017A&A...601A..95C}, and the fit derived in this study (solid black). Dashed grey lines indicate constant sSFR values. 
                        Additional data from \citet[][blue squares]{Curti2024} and \citet[][green stars]{Faisst2025} are also included.
                }
                \label{fig:MS_comparison}
        \end{figure}

        The SFRs were derived from Balmer emission lines (H$\alpha$ or H$\beta$). In particular, we converted the extinction-corrected H$\alpha$ luminosity, $L(\mathrm{H}\alpha)$, into the SFR using the calibration from \citet{Shapley_2023}. For the sake of consistency, all the SFRs from the literature were corrected to match the calibration of \cite{Shapley_2023}. This calibration is appropriate for subsolar metallicities. For galaxies without H$\alpha$ observations, we use the dust-corrected H$\beta$ flux scaled by the theoretical case B recombination factor (2.79) for $T_{\rm{e}}=15{,}000$~K and $n_{\rm{e}}=100~\mathrm{cm}^{-3}$ \citep{Perez-montero2017PASP..129d3001P}. The derived log(SFR / $M_\odot\,\mathrm{yr}^{-1}$) range from -0.24 to 2.38, with the 5th and 95th percentiles at 0.23 and 1.82. Combining these SFRs with our stellar mass estimations allows us to place the galaxies on the stellar mass-SFR plane and compare with the star-forming main sequence for galaxies at z $\sim$ 4-6 \citep{speagle2014,Calabro2024A&A...690A.290C}, as shown in Fig. \ref{fig:MS_comparison}. \\
        
        We fit the main sequence accounting for uncertainties in both axes (orange data in fig.\ref{fig:MS_comparison}), yielding a slope of $\alpha = 1.082$ and an intercept of $b =-8.730$, indicating a steeper relation than previously reported. For comparison, \citet{speagle2014} predict $\alpha\sim0.81$ at $z\sim5$, while \citet{Calabro2024A&A...690A.290C} find $\alpha\sim0.63$ for $4<z<6$. We obtained a steeper slope ($\alpha>1$) and relatively constant in specific star formation rate (sSFR) of $\sim$ 10$^{-8}$ yr$^{-1}$. We include in blue the point from \cite{Curti2024}, who reported median values of $\langle \mathrm{log(SFR/M_\odot\,\mathrm{yr}^{-1})} \rangle = 0.41$ and $\langle \log M_\star/M_{\odot} \rangle = 7.84$, while \cite{Nakajima2023ApJS..269...33N} provides $\langle \mathrm{log(SFR /M_\odot\,\mathrm{yr}^{-1}} \rangle = 0.69$ and $\langle \log M_\star /M_{\odot} \rangle = 8.67$. In comparison, our sample exhibits median values of $\langle \mathrm{log(SFR /M_\odot\,\mathrm{yr}^{-1}} \rangle = 0.97$ and $\langle \log M_\star/M_{\odot} \rangle = 9.15$. For completeness, we also consider the sample analysed by \citet{Faisst2025}, which targets a higher-mass regime, with median values of $\langle \mathrm{log(SFR /M_\odot\,\mathrm{yr}^{-1}} \rangle = 1.64$ and $\langle \log M_\star/M_{\odot} \rangle = 10.04$. Since our sample is the only one selected via the $\lambda$4363 line, we did not mix it with other datasets for the fit. In addition, we include a colourbar indicating the redshift of our galaxies. It shows a shift in normalisation with redshift, although the overall slope remains unchanged. At higher redshifts, the selection naturally favours galaxies that occupy the most extreme parts of the diagrams, generally corresponding to systems with higher specific star‑formation rates \citep[e.g.][]{speagle2014,Whitaker2014ApJ...795..104W,santini2017ApJ...847...76S}.
        
        \subsection{Metallicity}
        
\begin{figure}
                \centering
                \includegraphics[width=1\linewidth]{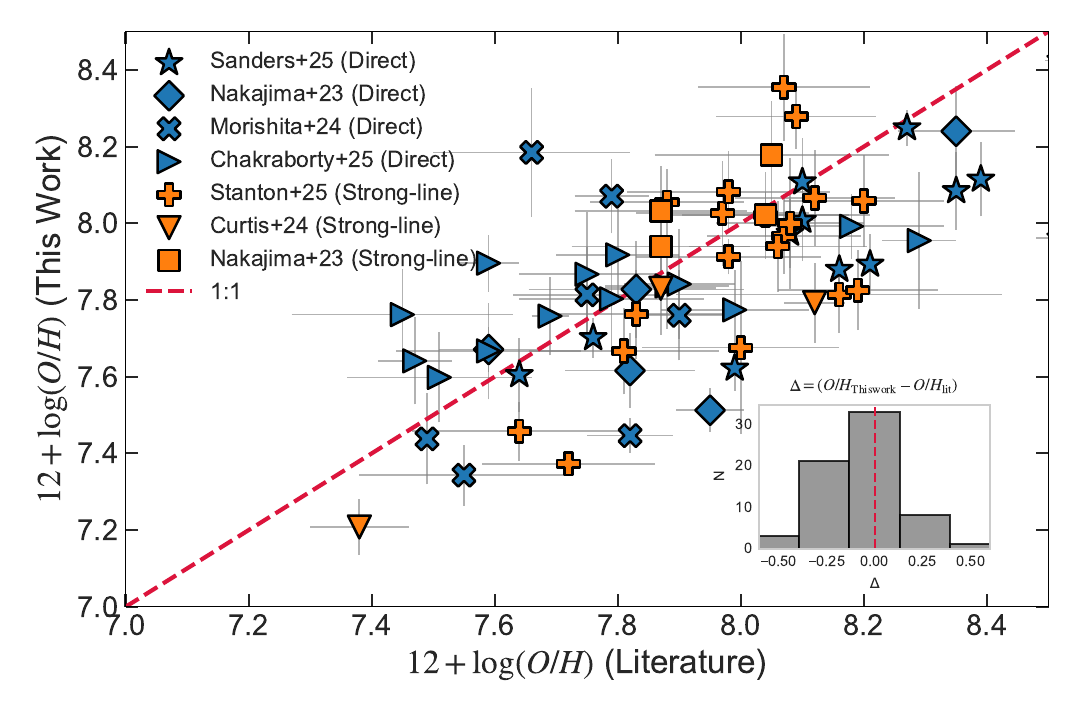}
                \caption{Comparison of oxygen abundances derived in this work with values reported in the literature. 
                        \textit{Main panel}: $12+\log(\mathrm{O/H})$ from this work versus literature values, 
                        with error bars representing measurement uncertainties. Direct-method abundances are shown in blue symbols \citep{Nakajima2023ApJS..269...33N,Morishita2024ApJ...963....9M,Chakraborty2025,Sanders2025arXiv250810099S}, while strong-line abundances are in orange \citep{Curti2024,Nakajima2023ApJS..269...33N,Stanton2025arXiv251100705S}.
                        The dashed red line indicates the 1:1 relation. 
                        \textit{Inset}: Histogram of differences $\Delta = (O/H_{\rm This\,Work} - O/H_{\rm lit})$, highlighting the distribution of residuals.}
                \label{fig:O/H comparatio}
        \end{figure}

        Next, we examined the chemical abundance analysis and validated our methodology against values reported in the literature. Figure~\ref{fig:O/H comparatio} compares our measurements with studies based on the direct method \citep{Nakajima2023ApJS..269...33N,Morishita2024ApJ...963....9M,Sanders2025arXiv250810099S,Chakraborty2025} and strong-line calibrations \citep{Nakajima2023ApJS..269...33N,Curti2024, Stanton2025arXiv251100705S}. For direct-method metallicities, the cross-matched sample includes 41 galaxies, with a mean residual of +$0.11$~dex. For strong-line metallicities, the comparison involves 29 sources, yielding a mean offset of +$0.22$~dex. This difference highlights the limitations of strong-line calibrations and reinforces the value of the direct method for reliable abundance determinations.\\
        
        A significant part of the discrepancy between our metallicity estimates and the direct-method values reported in the literature might arise from the treatment of the electron density and $T_{\rm{e}} (\mathrm{O\,II})$. At high redshift, auroral lines essential for determining $T_{\rm{e}} (\mathrm{O\,III})$ are often unavailable, requiring reliance on empirical prescriptions  \citep[e.g.][]{campbell10.1093/mnras/223.4.811, izotov2006A&A...448..955I, Hagele2006MNRAS.372..293H}. Assuming independence from electron density introduces a systematic shift, as noted in previous work (e.g. \citealt{Perez-montero2003MNRAS.346..105P, cataldi2025, Zinchenko2026}).
    
    To assess this possibility, we built a reference curve using the density-dependent $T_{\rm{e}} (\mathrm{O\,II})$--$T_{\rm{e}} (\mathrm{O\,III})$ relation of \citet{Hagele2006MNRAS.372..293H} at a fiducial density of $n_{\rm{e}}=300$~cm$^{-3}$, applied consistently both to estimate $T_{\rm{e}} (\mathrm{O\,II})$ and in the $n_{\rm{e}}$-dependent term of the O$^+$/H$^+$ equation. We compared this against four alternatives that isolate two separate effects. First, the density effect: we re-evaluated the same \citet{Hagele2006MNRAS.372..293H} relation at $n_{\rm{e}}=100$~cm$^{-3}$ instead of 300, again applying it consistently in both equations, which shows how much an incorrect density assumption alone biases the abundance. Second, the prescription effect: we replaced the density-dependent relation with three density-independent calibrations, the linear fit of \citet{campbell10.1093/mnras/223.4.811}, the polynomial fit of \citet{izotov2006A&A...448..955I}, and the linear fit of \citet{cataldi2025}, while keeping $n_{\rm{e}}=300$~cm$^{-3}$ fixed in the abundance equation, so that the only thing changing relative to the reference is the $T_{\rm{e}} (\mathrm{O\,II})$ prescription itself.
    
\begin{figure}
                \centering
                \includegraphics[width=1\linewidth]{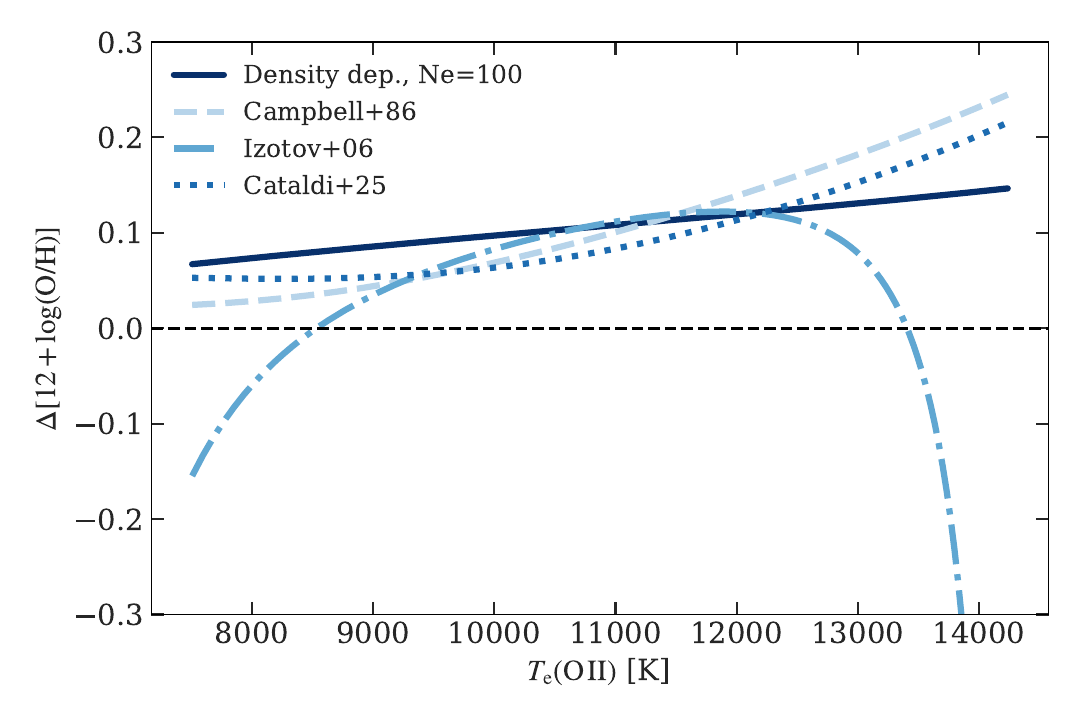}
                \caption{Difference in oxygen abundances as a function of the T[{O\,II}]. The reference curve is computed using a density-dependent T[{O\,II}] relation with $N_\mathrm{e} = 300~\mathrm{cm}^{-3}$. The four $\Delta$ curves show the difference relative to the reference when using T[{O\,II}]derived for $N_\mathrm{e} = 100~\mathrm{cm}^{-3}$ (density-dependent), the \cite{campbell10.1093/mnras/223.4.811}, \cite{izotov2006A&A...448..955I} and \citet{cataldi2025} prescriptions.}
                
                \label{fig:Delta_OH}
        \end{figure}

    Figure~\ref{fig:Delta_OH} shows $\Delta[12+\log(\mathrm{O/H})]$, the offset between the reference curve and each of these four cases, as a function of $T_{\rm{e}} (\mathrm{O\,II})$. All four overestimate the oxygen abundance relative to the reference across most of the temperature range, by 0.1--0.2~dex at $T_{\rm{e}} (\mathrm{O\,II})>10\,000$~K. The density effect alone (solid dark curve) grows smoothly with temperature. The \citet{campbell10.1093/mnras/223.4.811} and \citet{cataldi2025} relations behave similarly, since both scale close to linearly with $T_{\rm{e}} (\mathrm{O\,III})$. The \citet{izotov2006A&A...448..955I} polynomial relation agrees well with the others up to $T_{\rm{e}} (\mathrm{O\,III})\sim20\,000$~K, but turns over and drops at higher temperatures because its quadratic term takes over once the fit is extrapolated beyond the range it was calibrated for, which explains the sharp downturn seen in that curve. Overall, both the assumed density and the choice of $T_{\rm{e}} (\mathrm{O\,II})$ prescription bias the inferred oxygen abundance in the same direction, overestimating metallicity in the high-temperature, high-excitation regime typical of extreme emission line galaxies (EELGs).

\section{The mass-metallicity relation}
\label{mzr}

Using our sample of galaxies with robust detections of the [\ion{O\,III}] $\lambda4363$ auroral line, we investigated the shape of the MZR and its dependence on SFR. The use of direct, electron-temperature-based metallicities allows us to probe the low-mass, low-metallicity regime using a homogeneous dataset uniformly reduced and with all quantities derived in a consistent methodology. We compared our results with previous studies in the literature, highlighting both agreements and tensions, and exploring potential selection biases associated with $\lambda4363$-based samples that may influence the inferred MZR slope, normalisation, and its secondary dependence on SFR.\\
        
\begin{figure}
    \centering
    \includegraphics[width=1\linewidth]{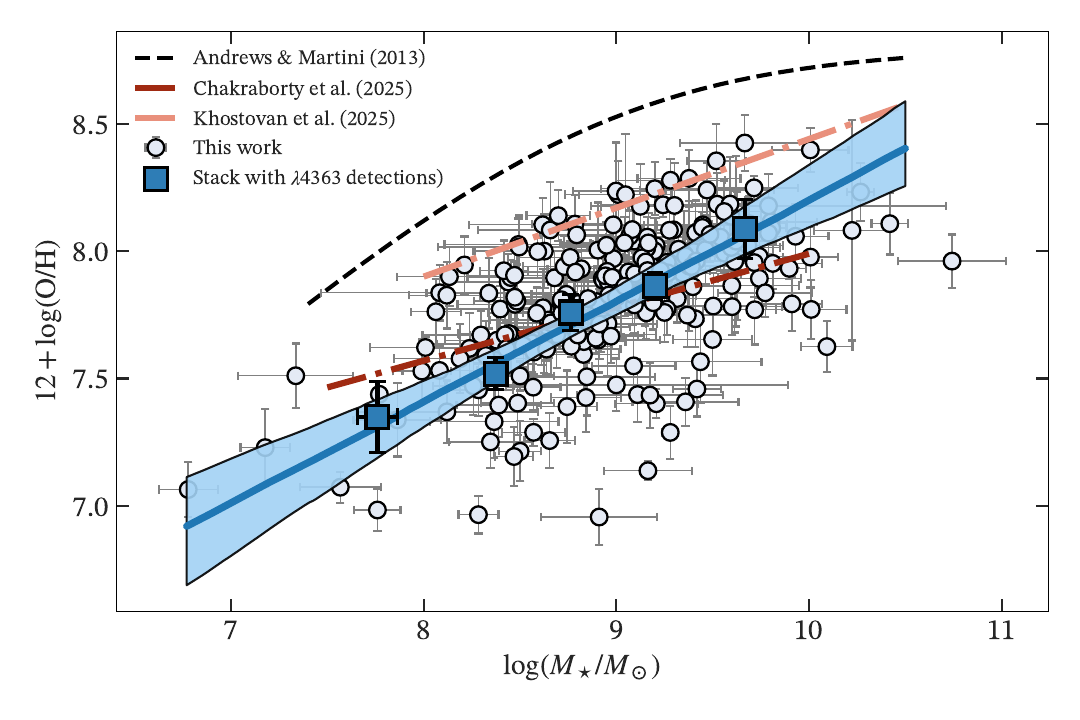}
    \caption{
        Mass-metallicity relation for our sample. Individual galaxies are shown as circles with error bars, while square symbols indicate metallicities computed in stellar-mass bins from the stacked spectra. The solid blue line represents the best-fitting linear MZR derived, with the shaded region showing the $1\sigma$ uncertainty. The dashed black curve corresponds to \cite{Andrews2013} (local universe relation). The red line corresponds to \cite{Chakraborty2025} and \citet{Khostovan2025} in pink.}
    
    \label{fig:MZR_direct}
\end{figure}

Figure~\ref{fig:MZR_direct} shows a strong monotonic relationship between stellar mass and oxygen abundance in our sample. To reduce the impact of intrinsic scatter and measurement uncertainties, we stack the data in log stellar mass bins $\log(M_\star/M_\odot)$: [7–8], [8–8.5], [8.5–9], [9–9.5], and [9.5–10]. The procedure is described in Appendix~\ref{sec.stacked}. We fit a linear MZR to the stacked points of the form,

\begin{equation}
    12+\log(\mathrm{O/H}) = \gamma\,(\log M_\star/M_{\odot} - 10) + Z_0.
    \label{eq_8}
\end{equation}

We fit the relation using \texttt{emcee} \citep{emcee2013PASP..125..306F},
accounting for uncertainties in both variables, and find
$\gamma = 0.38 \pm 0.09$ and $Z_0 = 8.19 \pm 0.13$.
We verified that this result is not driven by the choice of S/N threshold. Repeating the stacking analysis using only galaxies with
S/N $> 3$ in [O\,III]$\lambda 4363$ and refitting Eq.~\ref{eq_8} yields
$\gamma = 0.35 \pm 0.09$ and $Z_0 = 8.17 \pm 0.13$, in agreement with our
fiducial values well within $1\sigma$.

\begin{table*}[h!]
\centering
\small
\caption{Median properties of the stacks with (golden) and without (silver) [\ion{O\,III}] $\lambda4363$ detection.}
\resizebox{\textwidth}{!}{%
\begin{tabular}{c cc cc cc cc cc}
\hline
& \multicolumn{2}{c}{log (M$_\star$/M$_\odot$)} 
& \multicolumn{2}{c}{12+log(O/H)}
& \multicolumn{2}{c}{log(SFR M$_{\odot}$ yr$^{-1}$)}
& \multicolumn{2}{c}{EW(H$\beta$)\,\AA}
& \multicolumn{2}{c}{EW([O III])\,\AA} \\
Bin range
& Golden & Silver
& Golden & Silver
& Golden & Silver
& Golden & Silver
& Golden & Silver\\
\hline
$7.0 - 8.0$   
& $7.76 \pm 0.10$ & $7.79 \pm 0.05$
& $7.35 \pm 0.14$ & $7.75 \pm 0.11$
& $0.90 \pm 0.26$ & $0.28 \pm 0.165$
& $186\pm 175$ & $77 \pm 71$
& $800 \pm 639$ & $476 \pm 430$ \\

$8.0 - 8.5$   
& $8.36 \pm 0.02$ & $8.28 \pm 0.02$
& $7.51 \pm 0.06$ & $7.82 \pm 0.15$
& $0.73 \pm 0.07$ & $0.453 \pm 0.07$
& $107 \pm 36$ & $80 \pm 21$
& $680 \pm 212$ & $582 \pm 170$ \\

$8.5 - 9.0$   
& $8.78 \pm 0.02$ & $8.80 \pm 0.02$
& $7.75 \pm 0.16$ & $8.19 \pm 0.17$
& $0.88 \pm 0.048$ & $0.712 \pm 0.06$
& $98 \pm 19$ & $42 \pm 7$
& $651 \pm 119$ & $317 \pm 70$ \\

$9.0 - 9.5$   
& $9.20 \pm 0.02$ & $9.24 \pm 0.01$
& $7.86 \pm 0.09$ & $8.21 \pm 0.11$
& $1.06 \pm 0.05$ & $0.944 \pm 0.04$
& $88 \pm 11$ & $37 \pm 3 $
& $611 \pm 82$ & $219 \pm 21$ \\

$9.5 - 10.0$  
& $9.66 \pm 0.02$ & $9.68 \pm 0.02$
& $8.08 \pm 0.14$ & $8.28 \pm 0.12$
& $1.30 \pm 0.07$ & $1.30 \pm 0.05$
& $81 \pm 7$ & $31 \pm 2$
& $456 \pm 75$ & $122 \pm 9$ \\

\hline
\end{tabular}
}
\footnotesize Number of objects per bin from top to bottom: 7, 42, 71, 78 for the golden and 10, 50, 75, 176, 65 for the silver samples, respectively. Stellar mass and SFR are measured as the median of the galaxies within a bin. All the resulting stacks have [\ion{O\,III}] $\lambda4363$. 

\label{tab:bins_stats_binned_stacked}

\end{table*}

In the literature, recent JWST studies have reported a wide range of slopes and normalisations for the MZR. In Fig.~\ref{fig:MZR_direct}, we focus our comparisons on those using Te-based metallicities. In a first study, \citet{Chakraborty2025} presented the first JWST-based MZR at high redshift using exclusively direct-method measurements. Their sample, however, is significantly smaller (67 galaxies compared to our 242) and partly relies on lower-resolution PRISM/CLEAR spectroscopy, whereas our analysis is based on higher-resolution grating observations. In addition, they adopt a fixed electron density of $n_{\rm{e}} = 567\,\mathrm{cm^{-3}}$, higher than the median value (390 cm$^{-3}$) inferred in our work. They derive a slope of $\gamma = 0.21 \pm 0.12$ and a normalisation of $Z_{0} = 7.99 \pm 0.21$, based on three stellar-mass bins. More recently, \citet{Khostovan2025} used a sample of 34 galaxies from the AURORA survey, combined detections of [O\,III]~$\lambda4363$ and, when available, [O\,II]~$\lambda\lambda7320,7331$, and adopted electron densities either from the [S\,II] doublet or from a redshift-dependent prescription. They report a slope of $\gamma = 0.27 \pm 0.04$ and a higher normalisation of $Z_{0} = 8.44 \pm 0.04$, and find that galaxies with [O\,II] auroral detections tend to occupy higher masses and metallicities.\\

\begin{figure}
    \centering
    \includegraphics[width=1\linewidth]{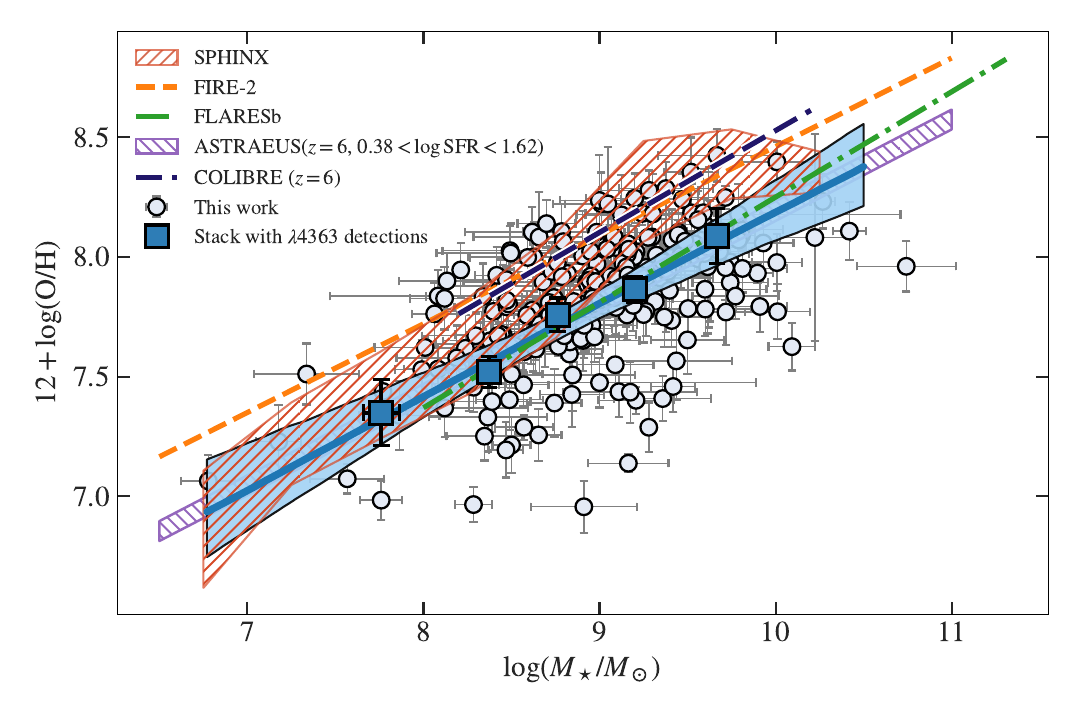}
    \caption{
        Black circles show individual JWST/NIRSpec galaxies with detections of the [{O\,III}] $\lambda4363$ auroral line from this work, while blue squares correspond to stacked measurements of galaxies with individual $\lambda4363$ detections. The solid blue line and shaded band indicate the median MZR and its intrinsic scatter inferred from our fitting. The hatched red region shows the prediction by the SPHINX simulations \citep{Katz2023OJAp....6E..44K}, while the dashed orange line corresponds to the FIRE-2 \citep{Marszewski2024ApJ...967L..41M}. The dashed-dotted green line shows the MZR from the FLARES simulations \citep{Lovell2021MNRAS.500.2127L,Vijayan2021MNRAS.501.3289V}. The purple hatched region illustrates the relation of \citet{cueto2024} $z=6$ over $0.38 < \log(\mathrm{SFR}/M_\odot\,\mathrm{yr}^{-1}) < 1.62$. The black dashed line shows the fit to the COLIBRE simulations at $z=6$ \citep{Sharda2026}.
}
    
    \label{fig:MZR_simul}
\end{figure}

In Fig.\ref{fig:MZR_simul}, we compare our measurements with predictions from cosmological simulations and models. Overall, most models populate a similar region of parameter space as our observations, indicating that current simulations broadly reproduce the average normalisation of the MZR at these redshifts. However, differences among the simulations reflect the underlying treatment of stellar feedback and star-formation variability. For example, FIRE-2 simulations \citep{Muratov2015MNRAS.454.2691M,Hopkins2018MNRAS.480..800H,Marszewski2024ApJ...967L..41M}, which include explicit modelling of stellar feedback and produce highly bursty star-formation histories, tend to predict higher metallicities at fixed stellar mass. In contrast, models such as FLARES \citep{Lovell2021MNRAS.500.2127L,Vijayan2021MNRAS.501.3289V}, based on smoother sub-grid prescriptions, exhibit a tighter and less stochastic relation with a slope in excellent agreement with our data. Similarly, ASTRAEUS \citep{cueto2024} also provides a closer match to our observations, reproducing both the normalisation and slope of the MZR across the explored stellar mass range, although without capturing the actual scatter. SPHINX \citep{Rosdahl2018MNRAS.479..994R, Katz2023OJAp....6E..44K}, designed to study galaxies during reionisation, occupies a similar mass regime but shows a different normalisation, likely reflecting differences in feedback strength, gas accretion histories, and resolution. Finally, the COLIBRE simulations \citep{Sharda2026} follow the multiphase ISM down to $\sim 10$~K with non-equilibrium H/He chemistry and a live dust model. During the epoch of reionisation, the best-fit $\gamma\approx$ 0.4 is in good agreement with our slope, but a normalisation $0.3$~dex above ours places it close to SPHINX.\\

The relatively large scatter at fixed mass and steep slope observed in the MZR from our data are qualitatively consistent with scenarios in which bursty star formation and episodic gas accretion play a significant role. Overall, the comparison suggests that models incorporating strong, time-dependent feedback and burst-driven evolution better match the diversity of galaxy properties observed in our sample. However, the models for which such a scatter is reported, such as SPHINX, do not recover the intrinsic scatter; nor can they account for the galaxies lying below the median relation at a fixed stellar mass, missing the most metal-poor galaxies for each stellar mass.

\subsection{Impact of auroral line selection on the shape of MZR}

The MZR presented in this work is derived from a galaxy sample explicitly selected by the detection of the [{O\,III}]$\lambda4363$ auroral line. While this requirement enables robust gas-phase metallicity measurements via the direct method, it also introduces selection effects that must be carefully considered. 

The detectability of the [{O\,III}]$\lambda4363$ line depends sensitively on the electron temperature of the ionised gas and therefore on metallicity. At fixed stellar mass, galaxies with higher oxygen abundances exhibit lower electron temperatures, causing the auroral line to rapidly weaken. As a result, samples selected through [{O\,III}]$\lambda4363$ detections are expected to be progressively biased towards low-metallicity systems over the mass range predominantly probed in this work ($6.77 \lesssim \log (M_\star/M_{\odot}) \lesssim 10.74$). To assess this bias, we performed a stacking analysis of galaxies without individual $\lambda$4363 detections. By stacking spectra within stellar mass bins, we recover the auroral line in the combined spectra and obtain direct metallicity measurements below the individual detection threshold.\\

In the first step, we repeated the selection of galaxies from the DJA catalogue, this time without imposing the S/N constraint on the [{O\,III}]$\lambda4363$ line. We followed all the previously described steps except for the diagnostic diagram, which was specifically designed for the auroral line. Instead, we employed alternative diagnostic diagrams: [{O\,III}]/H$\beta$ versus [Ne\,III]/[{O\,II}], following the methodology of \cite{Feuillet2024}; the classical \citet{bpt1981PASP...93....5B} BPT diagram, using the separation proposed by \cite{Kauffmann2003}; and He\,II /H$\beta$ versus [N\,II]/H$\alpha$, as defined in \cite{Shirazi2012}. As a result, the total number of selected galaxies is 1611. We then cross-matched this sample with the photometric catalogue and fitted their SEDs using the same procedure described in Sect. \ref{sed_fitting}. After this process, the final sample consists of 1241 galaxies. We then divided the sample into stellar mass bins. For each bin, we followed the same procedure employed for the 4363-detected sample, except that the redshift was determined using the H$\gamma$ line instead. The results are summarised in Table~\ref{tab:bins_stats_binned_stacked}. For brevity, we refer to the stacks constructed from galaxies with an [O\,III]$\lambda4363$ detection as the `golden sample' and those constructed from galaxies without a detection as the `silver sample'.

\begin{figure}
    \centering
    \includegraphics[width=1\linewidth]{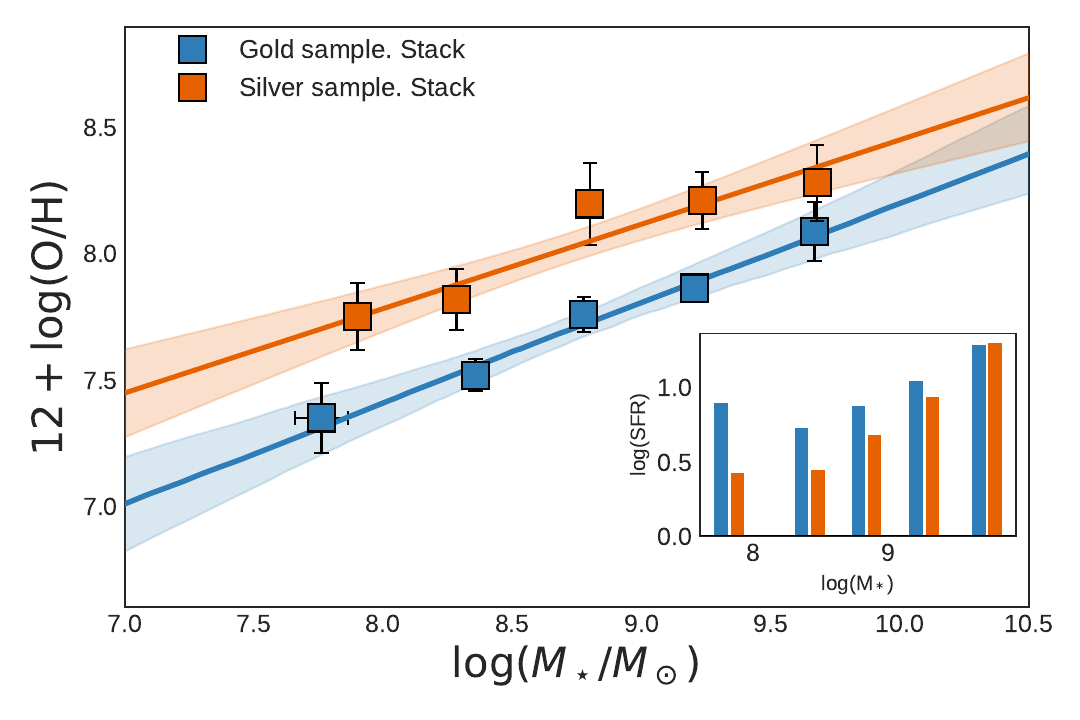}
\caption{MZR derived in this work.
Blue squares show metallicities measured in stellar mass bins using galaxies with a detected [O\,III]$\lambda4363$ line. Orange squares represent the metallicities of the stack done with galaxies with no [O\,III]$\lambda4363$ line in the spectra.
The solid lines represent the best-fit relation, while the shaded region indicates the corresponding $1\sigma$ uncertainty. The inset panel shows the median SFR as a function of stellar mass with the same colour pattern.
}
    \label{fig:MZr_stack}
\end{figure}

As shown in Fig.~\ref{fig:MZr_stack}, metallicities in the silver sample are systematically higher than those measured for stacks with [O\,III] $\lambda4363$ detections at fixed stellar mass, resulting in a higher normalisation of the MZR. The best-fitting parameters found are: $\gamma$=0.34$\pm$0.08
and Z$_0$=8.44$\pm$0.13. Our best-fit MZR for the silver sample is in good agreement with that reported in \citet{Khostovan2025}, with a very similar normalisation and slope. Their survey reaches greater depth, requiring the detection of auroral lines such as [O\,III]$\lambda$4363 and/or [O\,II] $\lambda\lambda$7320,7331. 

\subsection {Comparison with strong-line methods}
To compare the obtained results with MZR values from strong-line calibration studies, we applied the calibrations of \citet{Sanders2025arXiv250810099S} and \citet{Curti2024} to the golden and silver samples. These calibrations relate specific nebular emission-line ratios to the oxygen abundance, $12+\log(\mathrm{O/H})$, using polynomial parametrisations of metallicity-sensitive indices derived from best-fit relations to the $T_{\rm{e}}$-scale. We computed the following diagnostic ratios for each galaxy: $O3 = \mathrm{[O\,III]}\,\lambda 5007 / H\beta$, $O2 = \mathrm{[O\,II]}\,\lambda 3727 / H\beta$, $O32 = \mathrm{[O\,III]}\,\lambda 5007 / \mathrm{[O\,II]}\,\lambda 3727$,    $\hat{R}= 0.47 \log O2 + 0.88 \log O3$. In cases where multiple physical solutions exist (i.e. upper and lower branches), we employ O32 as an auxiliary line ratio to choose the appropriate branch. For each calibrator, we simultaneously fit all available line-ratio diagnostics by minimising a $\chi^2$ function. The $\chi^2$ includes both the propagated measurement uncertainties and the intrinsic scatter of each calibration, yielding a single best-fitting metallicity for each galaxy.\\

    \begin{figure}
        \centering
        \includegraphics[width=1\linewidth]{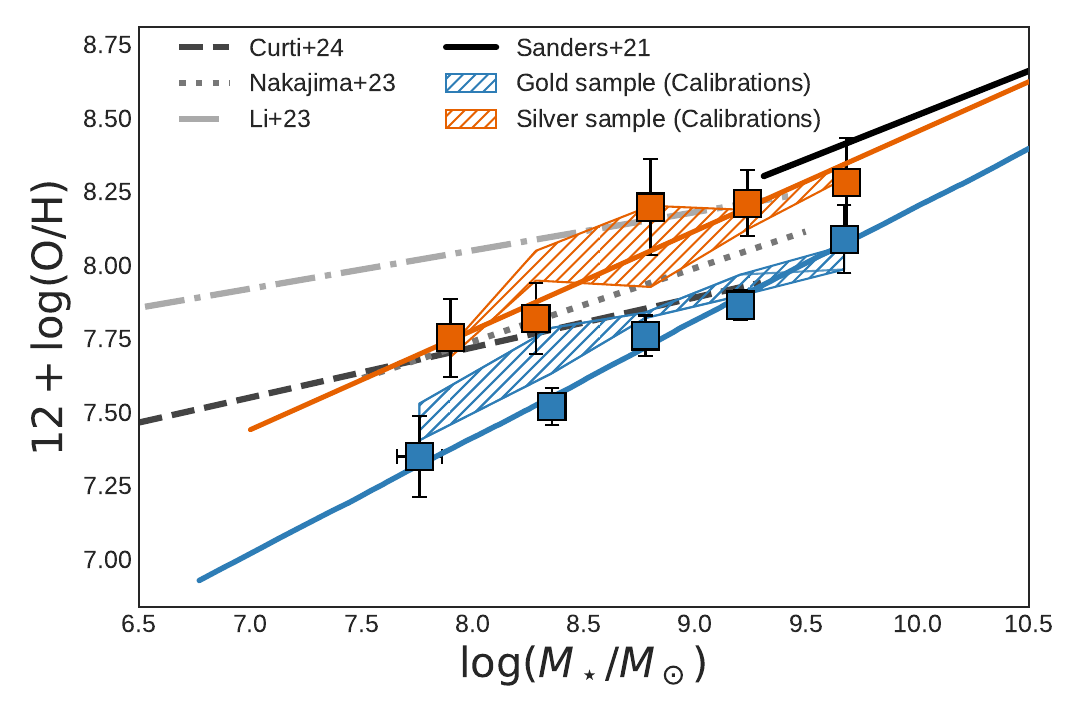}
        \caption{Comparison of the MZR obtained in this work with literature results based on strong-line methods. Blue squares show metallicities measured in stellar mass bins for galaxies with detected [O\,III]$\lambda$4363, while orange circles represent the metallicities measured from stacked spectra of galaxies without [O\,III]$\lambda$4363 detections. The shaded regions indicate the 1$\sigma$ uncertainties of the metallicities derived using the \cite{Sanders2025arXiv250810099S} and \cite{Curti2024} calibrations. The hatched regions mark the range spanned by the two calibrations for our golden and silver samples. Grey and black curves show MZR parametrisations from previous studies (\citealt{Curti2024}; \citealt{Nakajima2023ApJS..269...33N}; \citealt{Li2023}; \citealt{Sanders2021}).}
                \label{fig:SL_comparation}
    \end{figure}

In Fig.~\ref{fig:SL_comparation}, we present our results. The strong-line mass–metallicity relations for the golden and silver samples are shown by the blue and orange hatched regions, respectively, using the calibrations of \citet{Sanders2025arXiv250810099S} and \citet{Curti2024}. The shaded regions indicate the range of values spanned by these two calibrators. Our goal is not to assess which calibration performs better, but rather to highlight the deviations and dispersion introduced by adopting different calibrations. The calibrations exhibit discrepancies ranging from $\sim$0.05 to 0.25 dex across the stellar-mass range, but in some cases within the uncertainties. For the golden sample, the calibrations yield higher metallicities at lower stellar masses, while converging towards similar values at the high-mass end. In contrast, for the silver sample, the two methods show good agreement at low metallicities, but increasing discrepancies emerge at intermediate stellar masses. When comparing the MZR derived from strong-line calibrations in the literature (Table~\ref{tab:mzr_comparison}) with our results based on the direct method, we find that in the low-metallicity regime our golden sample yields a steeper slope ($\gamma = 0.38 \pm 0.09$) than that reported by \citet{Curti2024} ($\gamma = 0.17 \pm 0.03$). In contrast, the silver sample shows good agreement with the relation derived by \citet{Nakajima2023ApJS..269...33N} ($\gamma = 0.25 \pm 0.03$), although our fit yields a slightly steeper slope ($\gamma = 0.34 \pm 0.08$). The lensed galaxy sample analysed by \citet{Li2023} ($\gamma = 0.14 \pm 0.02$, $Z_0 = 8.31 \pm 0.06$) enables the exploration of intrinsically fainter systems and yields a higher normalisation at low metallicities compared to our direct-method. At the high-mass-and-high-metallicity end, the silver stack converges towards the relation reported by \citet{Sanders2021} ($\gamma = 0.30 \pm 0.02$, $Z_0 = 8.51 \pm 0.02$). \\

The individually detected $\lambda4363$ galaxies exhibit higher star-formation rates than those reported by \citet{Curti2024}, whereas the median SFR of the silver stack is lower and consistent with the typical values found in that study. This trend suggests that the differences in the MZR might be influenced by the differences in the SFHs.

\begin{table}[ht]
    \centering
    \small
    \caption{Linear MZR parameters from different studies, defined as in
    Eq.~\ref{eq_8}. }
    \label{tab:mzr_comparison}
    \begin{tabular}{@{}lcc@{}}
        \toprule
        Reference & $\gamma$ & $Z_0$ \\
        \midrule
        \multicolumn{3}{@{}l}{\textit{This work}} \\
        $\lambda4363$ detection & $0.38 \pm 0.09$ & $8.19 \pm 0.13$ \\
        $\lambda4363$ non-detection & $0.34 \pm 0.08$ & $8.44 \pm 0.11$ \\
        \addlinespace
        \hline
        \addlinespace
        \multicolumn{3}{@{}l}{\textit{Observations}} \\
        \protect\citet{Chakraborty2025} & $0.21 \pm 0.02$ & $7.99 \pm 0.21$ \\
        \protect\citet{Khostovan2025} & $0.27 \pm 0.04$ & $8.44 \pm 0.04$ \\
        \protect\citet{Curti2024} & $0.17 \pm 0.03$ & $8.06 \pm 0.18$ \\
        \protect\citet{Nakajima2023ApJS..269...33N} & $0.25 \pm 0.03$ & $8.24 \pm 0.05$ \\
        \protect\citet{Li2023} & $0.14 \pm 0.02$ & $8.31 \pm 0.06$ \\
        \protect\citet{Sanders2021} & $0.30 \pm 0.02$ & $8.51 \pm 0.02$ \\
        \protect\citet{Raptis2025} & $0.48 \pm 0.11$ & $8.58 \pm 0.14$ \\
        \addlinespace
        \hline
        \addlinespace
        \multicolumn{3}{@{}l}{\textit{Simulations and models}} \\
        ASTRAEUS \protect\citep{cueto2024} & $0.38 \pm 0.06$ & $8.21 \pm 0.04$ \\
        SPHINX \protect\citep{Katz2023OJAp....6E..44K} & $0.49 \pm 0.01$ & $8.53 \pm 0.02$ \\
        FIRE-2 \protect\citep{Marszewski2024ApJ...967L..41M} & $0.37$ & $8.46$ \\
        FLARES \protect\citep{Lovell2021MNRAS.500.2127L} & $0.44$ & $8.25$ \\
        COLIBRE \protect\citep{Sharda2026} & $0.43 \pm 0.01$ & $8.53 \pm 0.01$ \\
        \bottomrule
    \end{tabular}
    \tablefoot{SPHINX and FLARES values are taken from
    \citet{Katz2023OJAp....6E..44K} and \citet{Lovell2021MNRAS.500.2127L};
    see also \citet{Rosdahl2018MNRAS.479..994R} and
    \citet{Vijayan2021MNRAS.501.3289V}. Uncertainties quoted for the
    simulations correspond to the fits to the simulated relations and do not
    include systematic effects.}
\end{table}

\begin{figure*}
    \centering
    \includegraphics[width=1\linewidth]{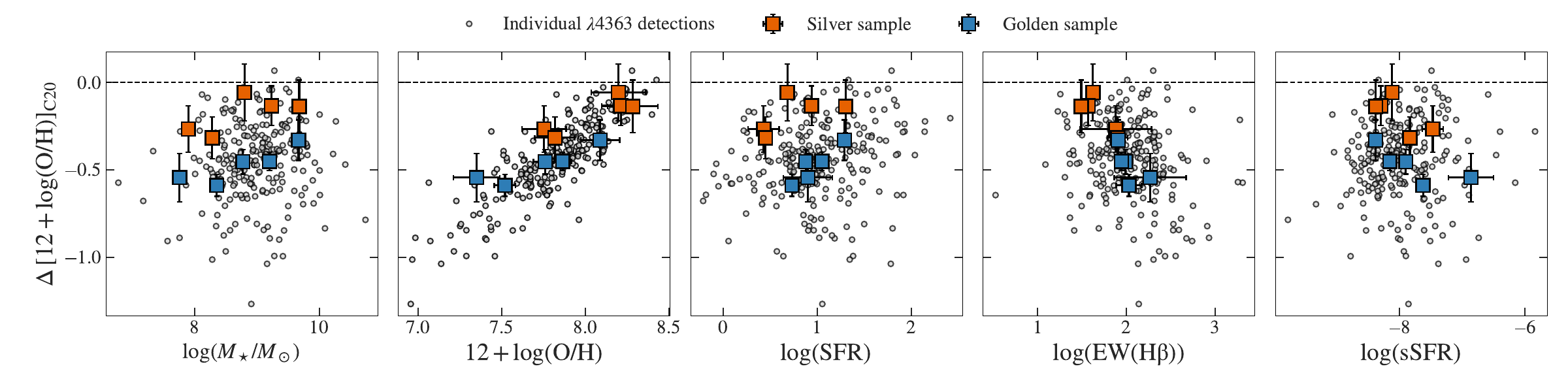}
    \caption{Residuals from the fundamental metallicity relation (FMR) of \citet{Curti2020}, expressed as $\Delta\, [12+\log(\mathrm{O/H})]_{\rm C20}$, shown as a function of stellar mass (left), gas-phase metallicity, SFR, EW(H$\beta$), and specific SFR (right). Grey points correspond to galaxies with detected [O III] $\lambda4363$, while the coloured markers show the stacked measurements for the silver (orange) and golden (blue) stacks. The horizontal dashed line marks $\Delta=0$.}
    \label{fig:FMR}
\end{figure*}
\subsection{Connection between EW SFH and the MZR}

The metallicity offset is accompanied by differences in star formation properties. At fixed stellar mass, the golden stack exhibits systematically higher SFRs than those inferred from the silver stack. For instance, in the $7.0 < \log M_\star/M_{\odot} < 8.0$ bin, the median SFR decreases from $\log(\mathrm{SFR/M_{\odot} yr^{-1}}) \simeq 0.9$  to $\log(\mathrm{SFR/M_{\odot }yr^{-1}}) \simeq 0.28$. This difference is larger for low masses. This effect was also shown by \citet{Marszewski2025ApJ...991L...4M}, who found that galaxies in the last SFR quintile of a mass-complete sample exhibit a higher MZR normalisation; however, this offset decreases towards higher stellar masses, where the relations converge. Further differences are observed when considering emission-line equivalent widths (EWs). At fixed stellar mass, the silver stack shows lower EW(H$\beta$) and EW([O\,III]) values compared to the golden stack. In the lowest-mass bin, EW(H$\beta$) decreases from $\sim$~186\,\AA\ to $\sim$77~\AA\, while EW([O\,III]) drops from 800\,\AA\ to 476\,\AA. This trend remains monotonic with stellar mass and is observed in both emission lines.\\

Taken together, these differences indicate that the silver stack spectra are dominated by galaxies with lower specific SFRs, older luminosity-weighted stellar populations, and less extreme ionisation conditions. In contrast, the golden preferentially traces galaxies with intense, recent star formation episodes, characterised by high SFRs and large emission-line EWs. The golden stacks, with high SFRs and large EWs, are consistent with a bursty star formation regime likely triggered by recent accretion of metal-poor gas. Such accretion simultaneously fuels star formation and dilutes the interstellar medium \citep[e.g.][]{rupke2010ApJ...723.1255R,borja2024NatAs}, resulting in lower metallicities at fixed stellar mass. Conversely, silver stacks appear to represent a more evolved population, in a more advanced stage of the burst, where star formation is reduced, gas temperature is lower, and chemical enrichment has progressed over longer timescales.\\

As a result, the observed MZR is best described as the superposition of at least two physically distinct sequences: a sequence traced by individually detected galaxies, with high SFRs, large EW(H$\beta$) and EW([O\,III]), and lower metallicities; and another revealed by stacking, characterised by lower SFRs, reduced EWs, and systematically higher metallicities at fixed stellar mass. This distinction may have important implications for the interpretation of the MZR, also when compared with low-redshift environments. A consistent comparison with low-redshift samples analysed using the same methodology is therefore required to study the change in shape of the MZR.\\

One more aspect that would favour this interpretation is the so-called fundamental metallicity relation (FMR; \citealt{Mannucci2010}). It links stellar mass, gas-phase metallicity, and star-formation rate. The FMR can be naturally interpreted within gas-regulator models, in which a balance between gas inflow, star formation, and outflows governs the evolution of galaxies. In this picture, the gas inflow rate approximately balances the sum of the star-formation rate and the outflow rate. Exploring how our data deviate from the FMR may therefore offer useful insight into the physical processes behind our observations. In Fig.\ref{fig:FMR} we show the quantity $\Delta$O/H computed using the FMR from \citet{Curti2020} and compare it with our results. At first glance, $\Delta$O/H is consistently negative, suggesting that our sample does not follow the FMR in \citet{Curti2020}, and it could imply significant deviations from predictions of gas regulator models \citep[e.g.][]{Dave2011,Lilly2013}. 

From the stacked (blue and orange points), we see a systematic trend: the golden stack lies below the silver stack. In other words, the stack built from galaxies without an [O\,III]$\lambda4363$ detection appears to lie closer to the FMR. We also find a strong positive correlation between $\Delta$O/H and O/H: the lower the metallicity, the larger the offset from the FMR \citep[also see][]{Laseter2025}. This trend suggests that low-metallicity galaxies, typically with a recent or ongoing episode of star formation, can drive them away from the physical conditions assumed by the FMR. Also, it has been demonstrated in the literature that a bursty SFH can reproduce the negative scatter found in the FMR \citep[e.g.][]{Pallottini2025A&A...699A...6P,Faisst2025}. On the other hand, \citet{McClymont2026MNRAS.tmp...20M} propose a scenario where FMR inverts in low-mass galaxies due to regular dilution by low-metallicity gas inflow without the need for a bursty SFH. \\

On a temporal scale, our results seem to suggest that the golden population traces the lower envelope of the MZR, representing systems with younger starbursts. Such episodes can push galaxies out of equilibrium \citep{Sharda2021MNRAS.504...53S,borja2024NatAs,kotiwale2026A&A...706A.165K}, driving them away from the FMR. As the star‑formation conditions settle and the activity reduces, feedback can regulate the gas reservoir more efficiently, allowing the galaxy to move back and drift towards the upper envelope of the MZR. The scatter between these two regimes likely reflects the diversity of star‑formation histories and the inflow mass rate that fuels recent star formation. 

\section{Chemical enrichment from nitrogen and helium abundances}

Besides oxygen, looking at elements such as nitrogen and helium gives us additional clues about how galaxies build up their metals over time. In this way, they can offer a more complete view of the processes that shape the chemical evolution of galaxies. In this section we first present the abundances obtained from nitrogen and helium, and finally, we discuss the results in the framework of this work. \\

\subsection{Nitrogen abundances}
Recent JWST observations have uncovered a population of high‑redshift galaxies showing nitrogen‑to‑oxygen ratios noticeably higher than those of local star‑forming systems at the same metallicity \citep[e.g.][]{finkelstain2022ApJ...940L..55F,Tang2023MNRAS.526.1657T,Isobe_2023,Castellano2024ApJ...972..143C,Schaerer2024A&A...687L..11S,Topping2024MNRAS.529.3301T,karla2025MNRAS.544.1588A,Morel2025arXiv251120484M,cameron2026arXiv260115964C}. Different mechanisms have been suggested to account for these elevated N/O values. From  the presence of massive stars or very massive stars, chemical enrichment of N/O via
WR stars, shocks, globular cluster precursors, contributions from fast-rotating massive Population III stars, even the presence of black holes \citep[e.g.][]{Senchyna2021MNRAS.503.6112S,Vink2023A&A...679L...9V,Cameron2023MNRAS.523.3516C,Kobayashi2024ApJ...962L...6K,Nandal2024A&A...688A.142N,Maiolino2024Natur.627...59M,Ji2026MNRAS.545f2110J,Schaerer2026A&A...708A..49S}. Although these scenarios differ in details, they share one ingredient: nitrogen must be produced on very short timescales, of the order of a few Myr, and often without a corresponding increase in oxygen. This points towards enrichment episodes occurring far from chemical equilibrium, together with inefficient metal mixing in the ISM of early galaxies.\\

Nitrogen exhibits a complex behaviour. At low metallicities, nitrogen is produced mainly as a primary element in massive stars, leading to an approximately constant N/O ratio (logN/O =~ -1.5). Only as metallicity approaches 12+log(O/H) $\sim$ 8.5, a mild secondary component begins to appear, causing N/O to rise slowly \cite[e.g.][]{Vila1993MNRAS.265..199V,perez-montero2009MNRAS.398..949P,Pilyugin2012MNRAS.421.1624P}. Nitrogen abundances were derived using the direct method, adopting 
T[O\,II] as representative of the low--ionisation zone. 
The ionic abundance of $\mathrm{N}^+$ was computed from the [NII] 
$\lambda6583$ emission line, including the contribution of the 
$\lambda6548$ line, assuming the theoretical ratio, 
$I_{6548} = I_{6583}/3$. Following \citet{Perez-montero2017PASP..129d3001P}, we have

\begin{align}
\frac{\mathrm{N}}{\mathrm{O}}
&=
\log\left(\frac{I_{6583}}{I_{3726}+I_{3729}}\right)
+ 0.493 \nonumber \\
&\quad
- 0.025\,t_{\mathrm{l}}
- \frac{0.687}{t_{\mathrm{l}}}
+ 0.1621\,\log t_{\mathrm{l}},
\end{align}

where, $t_{\mathrm{l}} \equiv \frac{T[O\,II]}{10^{4}\,\mathrm{K}}$.\\

\begin{figure}
    \centering

    \begin{subfigure}[t]{1\linewidth}
        \centering
        \includegraphics[width=\linewidth]{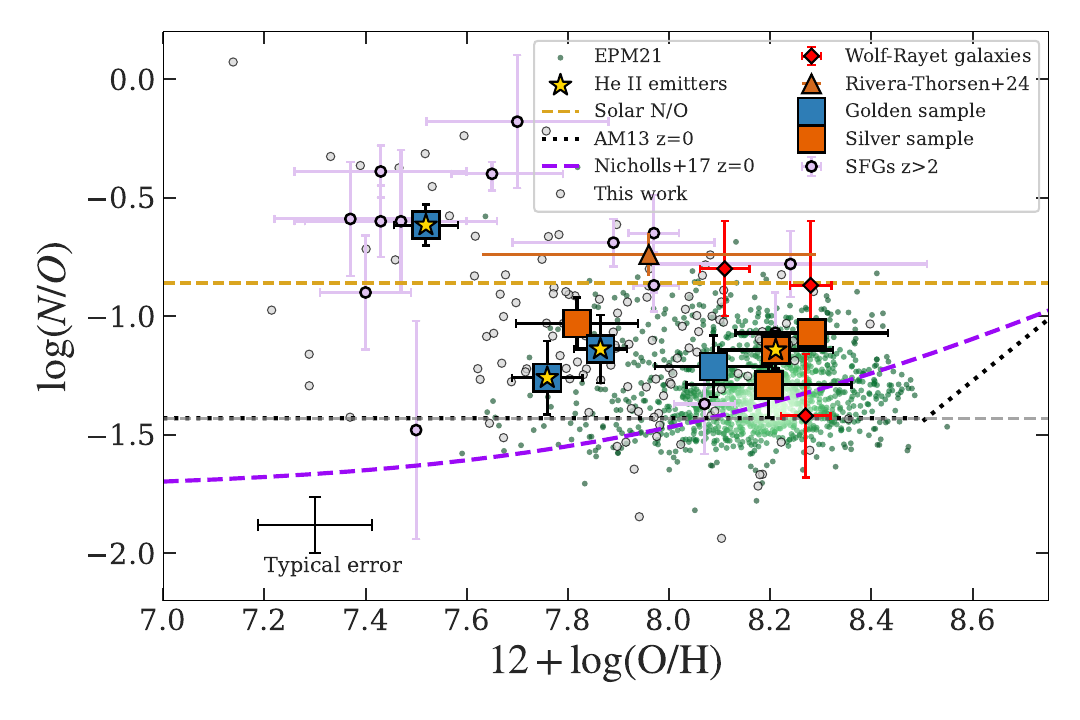}
        \caption{$N/O$ versus gas-phase metallicity.}
        \label{fig:NO_metallicity}

    \begin{subfigure}[t]{1\linewidth}
        \centering
        \includegraphics[width=\linewidth]{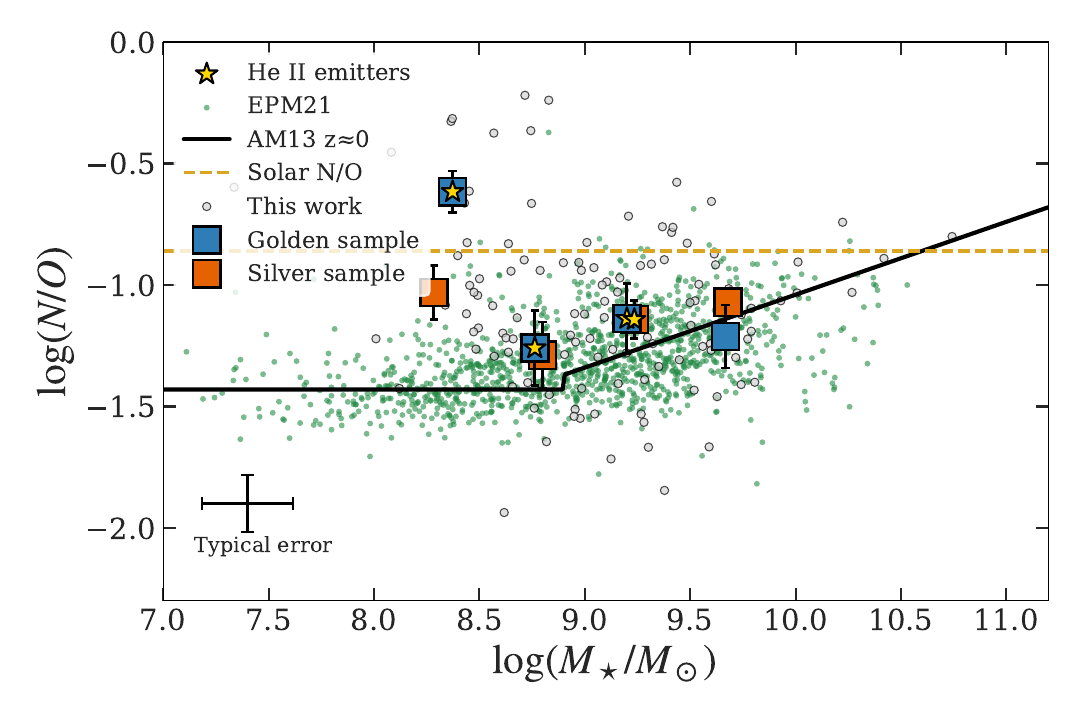}
        \caption{$N/O$ versus stellar mass.}
        \label{fig:NO_mass}
    \end{subfigure}
    
    \end{subfigure}

    \caption{
    Nitrogen-to-oxygen ratio for our sample shown against: (a) gas-phase metallicity and (b) stellar mass. Open circles indicate individual galaxies; coloured squares show the stacked measurements for the silver (orange) and golden (blue) subsamples. Error bars denote $1\sigma$ uncertainties. The solid blue curves indicate the $z \approx 0$ reference relations from Andrews \& Martini (2013). Yellow stars represent the stack where we have found He~II $\lambda$4686 emission. Green dots from \citet{enrique2021MNRAS.504.1237P} represent a sample of EELGs from SDSS. The purple curve is from \citet{Nichols10.1093/mnras/stw3235}. Light purple dots show the compilation of $z > 2$ nitrogen-enhanced galaxies measured with direct methods by \cite{arellano2025MNRAS.544.1588A}. Red squares indicate WR galaxies from \citet{vital2018MNRAS.478.5301F}. The brown triangle shows the measurement of the Sunburst arc (exhibiting WR features) from \citet{Rivera2024A&A...690A.269R}.
    }
    \label{fig:NO_panels}
\end{figure}

Figure~\ref{fig:NO_panels} shows how the nitrogen‑to‑oxygen ratio varies with stellar mass and gas‑phase metallicity. We do not include the lowest‑mass bin because no nitrogen emission lines were detected. In panel (a), we observe an enhancement of N/O relative to the primary plateau (N/O = -1.43 dex) obtained for metal-poor star‑forming dwarf galaxies and low‑metallicity HII regions in the Local Universe \citep{Andrews2013}. This enhancement is visible in both the Gold and Silver samples, which show similar values, except for one bin with 12+log(O/H)=7.51, which exhibits an even stronger enhancement relative to the others ($\log$(N/O)= -0.61). In green dots, we show the sample of local EELGs from \citet{enrique2021MNRAS.504.1237P}, where some galaxies populate the same region of parameter space. Significant He\,II $\lambda4686$ emission is detected in three out of the five stacks belonging to the golden sample and one in the silver sample, highlighted with a yellow star. All values can be found in table \ref{ratios}.\\

We also compare our measurements with the Sunburst Arc galaxy studied by \citet{Rivera2024A&A...690A.269R}, shown as a triangle, as well as with several nearby HII regions displaying clear WR spectral features (SHOC956, SHOC022, and SHOC955) from \citep{perez2013AdAst2013E..18P,vital2018MNRAS.478.5301F}. These objects occupy a similar region in metallicity compared to our galaxies, indicating that moderate nitrogen enhancement is observed in systems hosting WR populations.\\

\subsection{Helium abundance}
\begin{figure}
    \centering

    \begin{subfigure}[t]{1\linewidth}
        \centering
        \includegraphics[width=\linewidth]{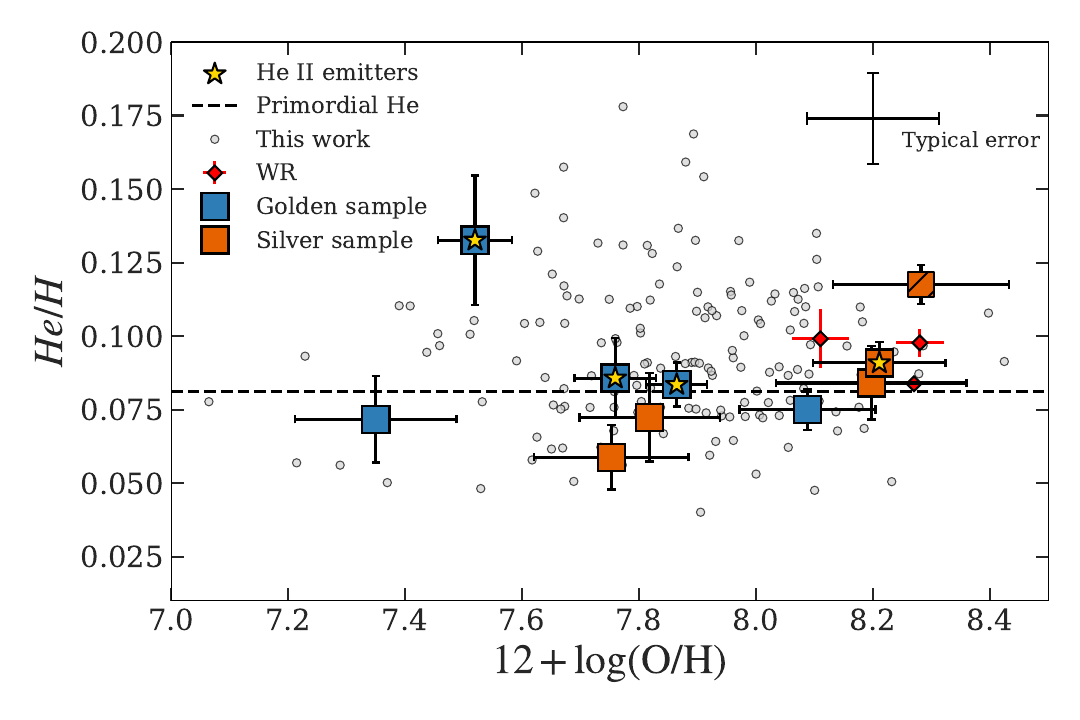}
        \caption{$He/H$ versus gas-phase metallicity.}
        \label{fig:he_metallicity}

    \begin{subfigure}[t]{1\linewidth}
        \centering
        \includegraphics[width=\linewidth]{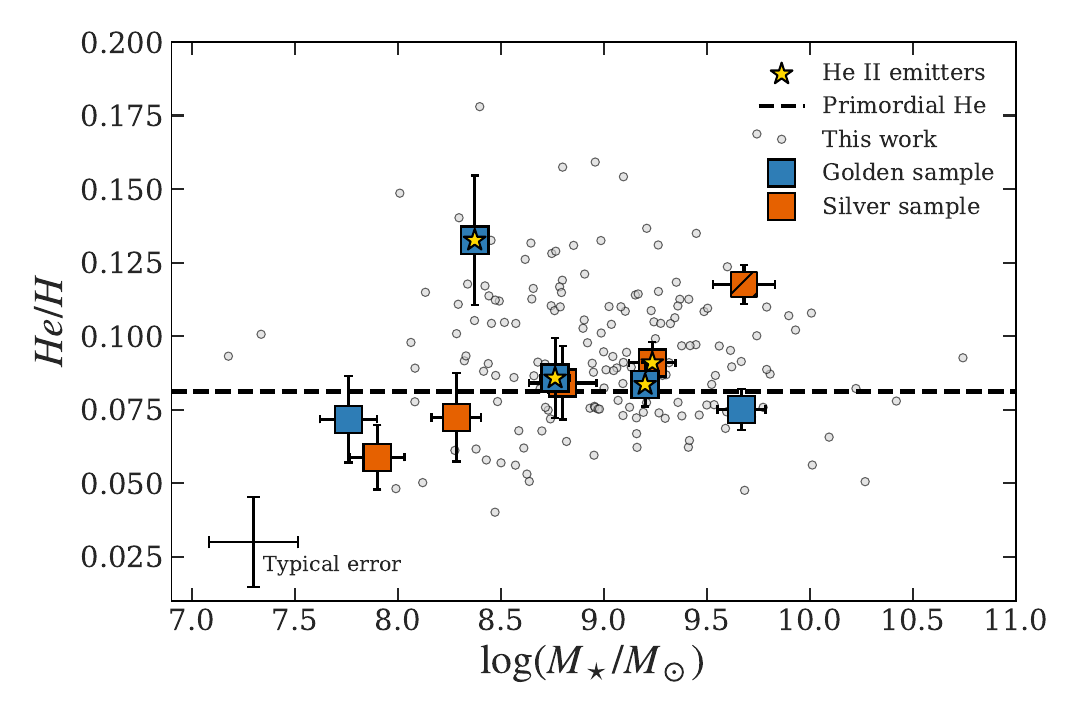}
        \caption{$He/H$ versus stellar mass.}
        \label{fig:he_mass}
    \end{subfigure}
    
    \end{subfigure}

    \caption{
    Helium-to-oxygen ratio for our sample shown against: (a) gas-phase metallicity and (b) stellar mass. 
    Open circles indicate individual galaxies; coloured squares show the stacked measurements for the 
    silver (orange) and golden (blue) subsamples. Error bars denote 1$\sigma$ uncertainties. Yellow stars represent the stack where we have found He\,II$\lambda$4686 emission. Squares crossed by a diagonal line indicate helium abundances requiring an ionisation correction larger than 15$\%$.
    }
    \label{fig:He_panels}
\end{figure}

Another important piece of information is the Helium abundance. Helium is mostly primordial, with an abundance in mass Y$_p$= 0.2453 or He/H = 0.0812 \cite[e.g][]{Peimbert2007ApJ,Planck}, with a modest contribution from stellar nucleosynthesis. There is general consensus on the evidence for enhanced helium and nitrogen abundances in the photospheres of most massive stars. Enhanced production of helium is expected during the life of massive stars, which eventually can expel it into the interstellar medium (e.g through strong mass loss, mass transfer in binaries, etc). 
The helium abundance is derived by accounting for the contributions from singly and doubly ionised helium, as well as a possible correction for neutral helium. The abundance of singly ionised helium, $\mathrm{He}^+/\mathrm{H}^+$, is measured using the HeI $\lambda5876$ recombination line. Line emissivities are computed with \textsc{PyNeb}, adopting the effective recombination coefficients of \citet{porter2012MNRAS.425L..28P,porter2013MNRAS.433L..89P}, which include corrections for collisional excitation. Electron temperatures and densities consistent with those derived from the nebular analysis are used in the emissivity calculations. The contribution from doubly ionised helium is estimated from the He\,II $\lambda4686$ recombination line, also using \textsc{PyNeb} emissivities. Significant He\,II emission is highlighted with a yellow star in Fig.~\ref{fig:He_panels}. For the remaining stacks, the He\,\textsc{ii} line is not detected, and its contribution is assumed to be negligible.\\

In addition to the ionised helium components, a fraction of neutral helium may be present within the ionised hydrogen regions. Given the high degree of ionisation of our sample, particularly for the golden sample, the neutral helium fraction is expected to be small, but not in the case of the silver one. In highly ionised H\,\textsc{ii} regions, the volumes occupied by $\mathrm{He}^+$ and $\mathrm{H}^+$ largely coincide, and the ionisation correction factor (ICF) for neutral helium approaches unity. However, the exact value of this correction depends on both the ionising spectrum and the gas distribution within the emitting region. A more accurate correction for neutral helium is obtained using the $\eta$ parameter, which relates the ionisation structure of helium and hydrogen through the ratios of successive ionisation stages of oxygen and sulphur \citep{vilchez1989Ap&SS.157....9V}. However, this approach requires reliable measurements of both $\mathrm{S}^+/\mathrm{S}^{++}$, which are not available for our sample. Instead, we follow the empirical prescriptions of \citet{peimbert1974RMxAA...1..129P} and \citet{kunth1983ApJ...273...81K}. We express the total helium abundance as
\begin{equation}
\frac{\mathrm{He}}{\mathrm{H}} = \mathrm{ICF}(\mathrm{He}) \left( \frac{\mathrm{He}^+ + \mathrm{He}^{++}}{\mathrm{H}^+} \right),
\end{equation}
with
\begin{equation}
\mathrm{ICF}(\mathrm{He}) = \left( 1 - 0.25\,\frac{\mathrm{O}^+}{\mathrm{O}} \right)^{-1}.
\end{equation}

Given the high‑ionisation conditions observed in most of our stacks, we adopt $\mathrm{ICF}(\mathrm{He}) = 1$ whenever He\,II is detected or when the ionisation parameter is sufficiently high, $\mathrm{O32} > 5$. Whenever the required correction exceeds 15$\%$ (ICF > 1.15, i.e. O$^+$/O > 0.52), we flag the resulting He/H abundance with a diagonal line to indicate that it is uncertain, since in this regime the total abundance depends strongly on the assumed ionisation structure.

\begin{table*}[h!]
\centering
\small
\caption{Stellar mass and metallicity ratios for galaxies with (golden) and without (silver) [\ion{O\,III}] $\lambda4363$ detection. }
\resizebox{\textwidth}{!}{%
\begin{tabular}{c cc cc cc cc}
\hline
& \multicolumn{2}{c}{log (M$_\star$/M$_{\odot}$)} 
& \multicolumn{2}{c}{12+log(O/H)}
& \multicolumn{2}{c}{log(N/O)}

& \multicolumn{2}{c}{He/H} \\
Bin range
& Golden & Silver
& Golden & Silver
& Golden & Silver
& Golden & Silver
\\
\hline
$7.0 - 8.0$   
& $7.76 \pm 0.10$ & $7.79 \pm 0.05$
& $7.35 \pm 0.14$ & $7.75 \pm 0.11$
& - & -

& $0.071 \pm 0.014$ & $0.058 \pm 0.011$ \\

$8.0 - 8.5$   
& $8.36 \pm 0.02$ & $8.28 \pm 0.02$
& $7.51 \pm 0.06$ & $7.82 \pm 0.15$
& $-0.62 \pm 0.08$ & $-1.03 \pm 0.11$

& $0.13 \pm 0.02$ & $0.072 \pm 0.015$ \\

$8.5 - 9.0$   
& $8.78 \pm 0.02$ & $8.80 \pm 0.02$
& $7.75 \pm 0.16$ & $8.19 \pm 0.17$
& $-1.26 \pm 0.15$ & $-1.29 \pm 0.14$

& $0.086 \pm 0.014$ & $0.074 \pm 0.012$ \\

$9.0 - 9.5$   
& $9.20 \pm 0.02$ & $9.24 \pm 0.01$
& $7.86 \pm 0.09$ & $8.21 \pm 0.11$
& $-1.14 \pm 0.14$ & $-1.14 \pm 0.08$

& $0.084 \pm 0.007$ & $0.09 \pm 0.007$ \\

$9.5 - 10.0$  
& $9.66 \pm 0.02$ & $9.68 \pm 0.02$
& $8.08 \pm 0.14$ & $8.28 \pm 0.12$
& $-1.21 \pm 0.13$ & $-1.07 \pm 0.06$

& $0.067 \pm 0.006$ & $0.097 \pm 0.007$ \\

\hline
\end{tabular}
}
\label{ratios}

\footnotesize
Number of objects per bin: 7, 42, 71, 78 for the Golden Sample and 10, 50, 75, 176, 65 for the Silver Sample.

\end{table*}

In Fig.~\ref{fig:He_panels}, we show the distribution of helium abundance as a function of stellar mass and metallicity. A clear enhancement in helium is observed for the golden sample in the stellar mass bin $8.0 \leq \log(M_\star/M_\odot) < 8.5$. This increase in helium abundance is consistent with the level of helium variation ($\sim$0.1 dex) observed in galaxies of similar mass, when moving from central regions to the outskirts \citep{Mendezdelgado2020MNRAS.496.2726M}. This suggests that the helium enhancement we derive in these galaxies may represent the equivalent of the stellar evolution and chemical enrichment processes in the entire life of a galaxy. Significant He\,II $\lambda4686$ emission is detected in three out of the five stacks belonging to the golden sample and to the silver sample, highlighted with a yellow star.

\subsection{What drives N and He enhancement and its connection with the MZR}

\begin{figure}
    \centering
    \includegraphics[width=1\linewidth]{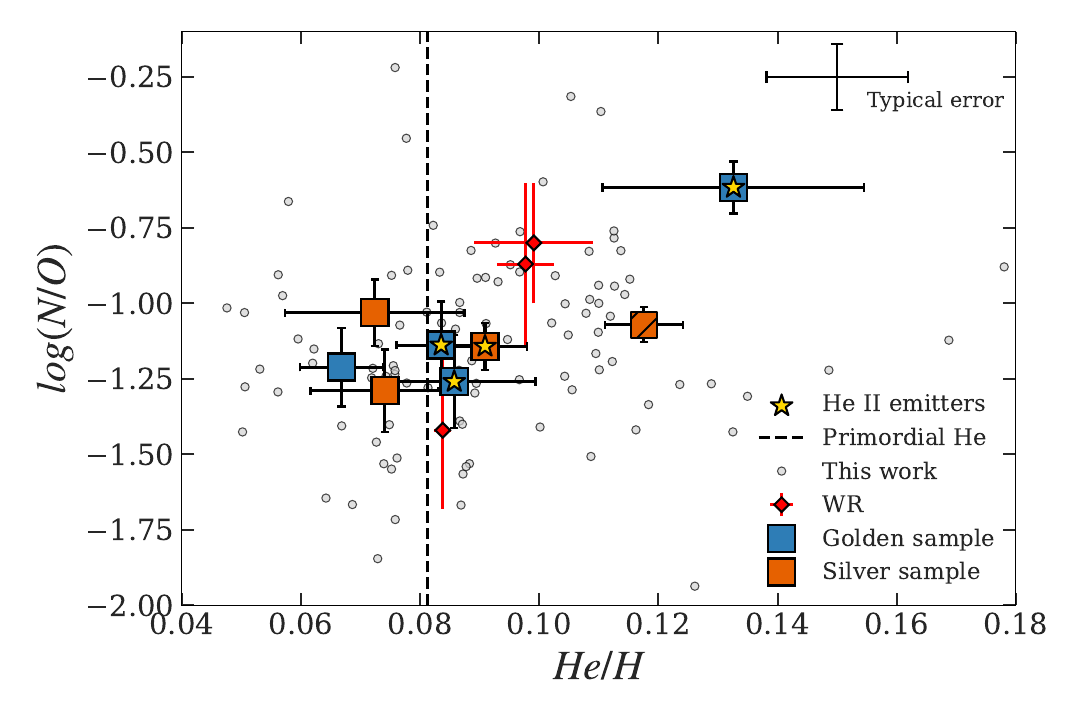}
    \caption{Nitrogen-to-oxygen ratio as a function of helium abundance for individual galaxies and stacked samples. Individual galaxies are shown as open circles, with a representative typical error indicated in the lower-left corner. Stacked measurements for the golden and silver subsamples are shown as blue and orange squares, respectively. Galaxies with significant He\,II$\lambda4686$ emission are highlighted with gold stars. The vertical dashed line marks the primordial helium abundance ($Y_\mathrm{p} = 0.2453$; He/H\,$= 0.0812$). The filled diamond indicates the WR comparison objects from the literature (see text). Squares crossed by a diagonal line indicate helium abundances requiring an ionisation correction larger than 15$\%$.}
    \label{fig:N/O_he}
\end{figure}

Additional insight is provided by the derived helium abundances \citep[e.g.][]{Ji2026MNRAS.545f2110J,Ebihara2026arXiv260104344E}. Figure~\ref{fig:N/O_he} shows the relation between helium and nitrogen abundance ratios, indicating that systems with enhanced N/O also tend to exhibit elevated He/H values. The bin with the strongest nitrogen enhancement also displays the highest helium abundance (He/H = 0.13). Interestingly, the stacked bins with significant He II emission tend to lie above the primordial helium abundance (dashed line), potentially linking hard ionising sources with enhanced helium abundances. For reference, nearby WR regions (shown in red) display helium abundances that overlap with those found in our most enriched bins, which may hint at a similar enrichment channel, although a larger sample would be needed to draw firmer conclusions.

\begin{figure}
    \centering
    \includegraphics[width=1\linewidth]{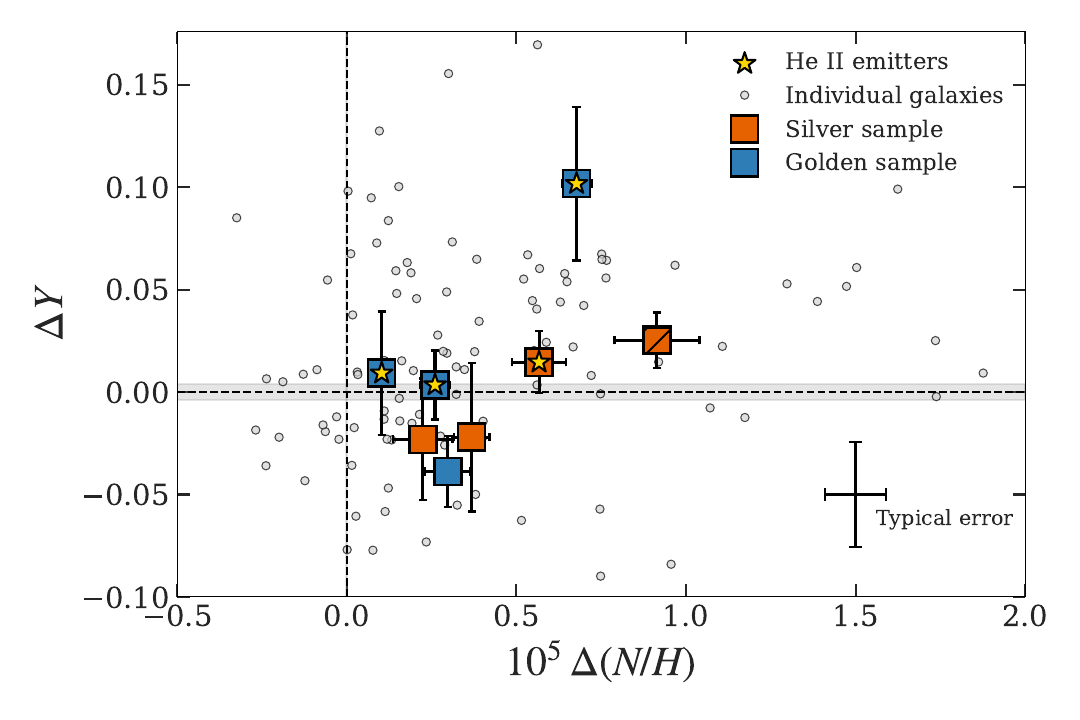}
    \caption{Relation between changes in helium abundance ($\Delta Y$) and nitrogen abundance ($\Delta N/H$) for individual galaxies and stacked samples. Individual galaxies are shown as light-grey circles, and the representative typical error is indicated in the lower-left corner. Silver and Golden samples are plotted as squares in orange and blue, respectively. He\,\textsc{ii}-emitting galaxies are highlighted with gold stars. Horizontal and vertical dashed lines indicate the reference $\Delta Y = 0$ and $\Delta N/H = 0$ levels. Shaded bands show the typical uncertainties. Squares crossed by a diagonal line indicate helium abundances requiring an ionisation correction larger than 15$\%$.
}
    \label{fig:noHE}
\end{figure}

Besides that, we attempted to disentangle the contributions of oxygen and nitrogen by computing the excess of helium ($\Delta \mathrm{He/H}$) versus the excess of nitrogen ($\Delta \mathrm{N/H}$). We assumed the empirical relation between helium and O/H for star-forming galaxies from \citet{izotov2004ApJ...602..200I} and adopted the theoretical N/O ratio plateau of $-1.43$ for our galaxies. This is shown in Fig.~\ref{fig:noHE}. We find that the largest helium enhancements correspond to relatively high nitrogen excesses. Additionally, we observe an increasing trend in the silver sample, consistent with the onset of secondary nitrogen production in the most massive galaxies. The coincidence of enhanced N/H and He/H in the same system is particularly suggestive in this direction, since both elements are expected to be released together through the winds of massive stars during their early stages of stellar evolution. Such enrichment could arise from WR stars, whose nitrogen-rich stellar winds appear within a few million years after the onset of star formation\footnote{Besides single star evolution considerations, introducing binarity could effectively increase the total mass transfer.}, or from very massive stars undergoing efficient rotational mixing, which can bring CNO-cycle products to the stellar surface early in their evolution. Indeed, \citet{Molla2012MNRAS.425.1696M} showed that stellar winds from WR stars can enhance N/H by up to $\sim1$ dex across a wide range of metallicities. Nevertheless, we cannot rule out an additional contribution from very massive stars, particularly in the system showing the highest enrichment. To better constrain the origin of these enhancements, we would also need to examine the C/O abundance \citep[e.g.][]{Berg2019,tapia2024MNRAS.534.2086T}.
This remains challenging at high redshift: the most robust diagnostic, C\,III]$\lambda 1908$/[O\,III]$\lambda 1666$, requires the much fainter $\lambda 1666$ line, while combining C\,III] with the optical [O\,III] lines makes the result strongly dependent on the dust correction. We will address this in forthcoming work.\\

\citet{cameron2026arXiv260115964C} suggested that these elevated abundances of N are caused by enrichment from young massive stars in extreme environments during extreme starbursts, and this seems to agree with the dual-burst scenario proposed by \citet{Kobayashi2024ApJ...962L...6K}. If we link this scenario with the information from the MZR, SFR, and EW, we find that our golden sample, which exhibits higher EW(H$\beta$) indicative of younger bursts and less chemically evolved systems, lies outside the equilibrium proposed by the FMR. These galaxies may be undergoing intense bursts with WR or supermassive stars (SMSs) capable of elevating the N/O ratio. When analysing the N/O ratio as a function of stellar mass, it is particularly remarkable that the most massive bins approach the onset of secondary nitrogen production via the CNO cycle. In contrast, considering the N/O–O/H relation, they have not yet reached the corresponding secondary regime. This apparent offset suggests that the oxygen abundances of these galaxies are lower than expected for their stellar masses, consistent with systems temporarily displaced from equilibrium by recent gas accretion that triggered a burst of star formation. A similar behaviour is observed in the sample of local analogues \citep[Fig.~\ref{fig:NO_metallicity}, green dots,][]{enrique2021MNRAS.504.1237P}, where some galaxies do the same transition in N/O-O/H and N/O-M$_{\odot}$. On the other hand, the silver sample is at a different evolutionary stage: although the SFR is still high, the bursts are older, they are slightly more chemically evolved, and we generally do not observe He II emission, features explained by WR stars or very massive stars, except in one bin. In these cases, the nitrogen enrichment appears consistent with natural galaxy evolution.\\

A promising way to test this scenario directly would be to resolve where the nitrogen and helium emissions originate. Spatially resolved integral-field spectroscopy could reveal whether N and He emission come from the same regions, as expected if both trace WR or very massive stars, or are spatially offset, which would instead point to distinct enrichment channels.

\section{Summary and conclusions}

In this work, we present one of the largest samples of star-forming galaxies with direct, electron-temperature–based metallicities spanning $0.8 \lesssim z \lesssim 9.1$, enabled by JWST/NIRSpec detections of the [O\,III] $\lambda4363$ auroral line. We used data from the DJA, which includes photometric catalogues and spectroscopic data. For the spectroscopic sample, we selected the latest public NIRSpec datasets v4, focusing only on sources observed with medium–high spectral resolution and excluding prism spectroscopy. This homogeneous approach allows us to probe the low-mass, low-metallicity regime of the MZR in a fully consistent way, with reduced systematic uncertainties. Our main findings can be summarised as follows:

\begin{enumerate}
    \item We derived an MZR for galaxies with individual [O\,III] $\lambda4363$ detections, spanning $\log(M_\star/M_\odot) \simeq 6.8$--$10.5$, using a direct $T_{\rm{e}}$-method. We found a slope of $\gamma = 0.38 \pm 0.09$ and $Z_0 = 8.19 \pm 0.13$, consistent with the range of values reported by recent studies at similar redshifts, albeit lying towards the steeper end. This behaviour is broadly consistent with hydrodynamical simulations and semi-analytic models.

    \item We recovered a Te-based MZR for galaxies without 4363 detections using stacking in bins of ($\log(M_\star/M_\odot)$): [7–8], [8–8.5], [8.5–9], [9–9.5], and [9.5–10]. The oxygen abundances derived from this stacked sample at fixed stellar mass are systematically higher ($\sim$0.3 dex) than those inferred from galaxies with individual [O\,III] $\lambda4363$ detections. The resulting MZR exhibits a similar slope ($\gamma = 0.34 \pm 0.08$), but with a higher normalisation of $Z_0 = 8.44 \pm 0.11$. This result suggests that samples selected via auroral-line detections preferentially trace the low-metallicity envelope of the MZR.

    \item At a given stellar mass, galaxies with detected [O\,III] $\lambda4363$ exhibit higher SFRs, larger emission line EWs, and larger offsets from the local fundamental metallicity relation. These properties are consistent with bursty star formation episodes and lower gas-phase metallicities. In contrast, galaxies without individual auroral-line detections appear more chemically evolved and closer to the local fundamental metallicity relation. These results suggest that the scatter and slope of the MZR in low-mass, high-redshift galaxies are linked to variations in their recent SFH through gas accretion and bursty evolution.

    \item Enhanced N/O ratios at low metallicity and indications of helium enrichment in some stacked spectra (log(N/O) = $-0.61 \pm 0.09$ and He/H $\sim$ 0.13) suggest the presence of rapid, short-timescale enrichment processes. These trends are qualitatively consistent with contributions from massive stellar populations or WR stars. Further constraints on the origin of these abundances will require additional measurements, such as the C/O ratio.
\end{enumerate}

Our analysis highlights the need to increase statistics at low stellar masses and to derive metallicities for galaxies that are not limited to strong starbursts (including systems with lower EWs and without detections of [O\,III] $\lambda4363$) to characterise the high-metallicity envelope of the MZR at low masses. Future comparisons with low-redshift samples analysed using consistent Te-methods will be essential to robustly trace the shape of the mass–metallicity relation across cosmic times. Observations of gravitationally lensed galaxies could extend the mass and metallicity ranges to lower values. Spatially resolved IFU observations would further clarify whether the nitrogen and helium emission are co-spatial, as expected for a WR or very massive star origin, or whether it is spatially offset, tracing distinct enrichment channels. Finally, further detailed chemical evolution models would be help improve our understanding of the different enrichment paths of these systems and to place appropriate constraints on their parameters.

\begin{acknowledgements}

We thank the anonymous referee for useful comments that contributed to improving the original manuscript. We acknowledge financial support from the Severo Ochoa grant CEX2021-001131-S, funded by MICIU/AEI/10.13039/501100011033. Also, we acknowledge FPI support under grant code CEX2021-001131-S-20-7 and support from the research grant PID2022-136598NB-C32 ("Estallidos8"). RA acknowledges support from Grant PID2023-147386NB-I00 funded by MICIU/AEI/10.13039/501100011033 and by ERDF/EU. We acknowledge the many developers and contributors to the JWST Science Calibration Pipeline, which we use for the initial reduction steps of both imaging and spectroscopic datasets. We thank Gabe Brammer and the DJA Team for their work in developing and maintaining the repository.
\end{acknowledgements}

        \bibliographystyle{bibtex/aa} 
        \bibliography{bibtex/references.bib}

\appendix
\onecolumn
\section{Diagnostic diagram for the stacks}

\begin{figure}[htbp]
    \centering
    \begin{subfigure}[t]{0.48\textwidth}
        \centering
        \includegraphics[width=\linewidth]{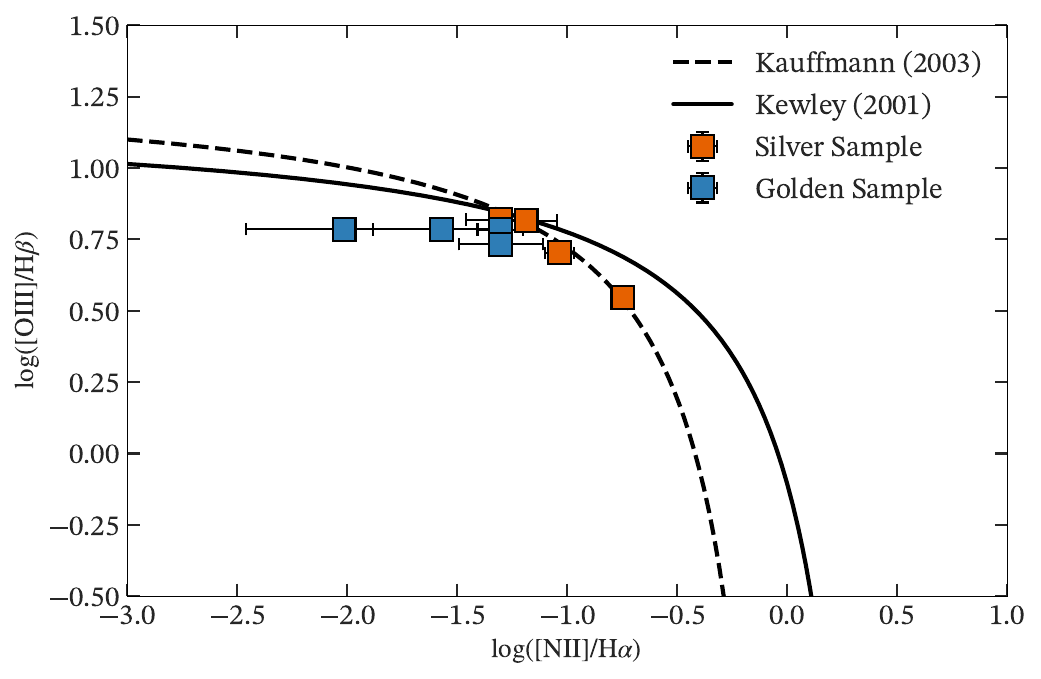}
        \label{fig:BPT}
    \end{subfigure}
    \hfill
    \begin{subfigure}[t]{0.48\textwidth}
        \centering
        \includegraphics[width=\linewidth]{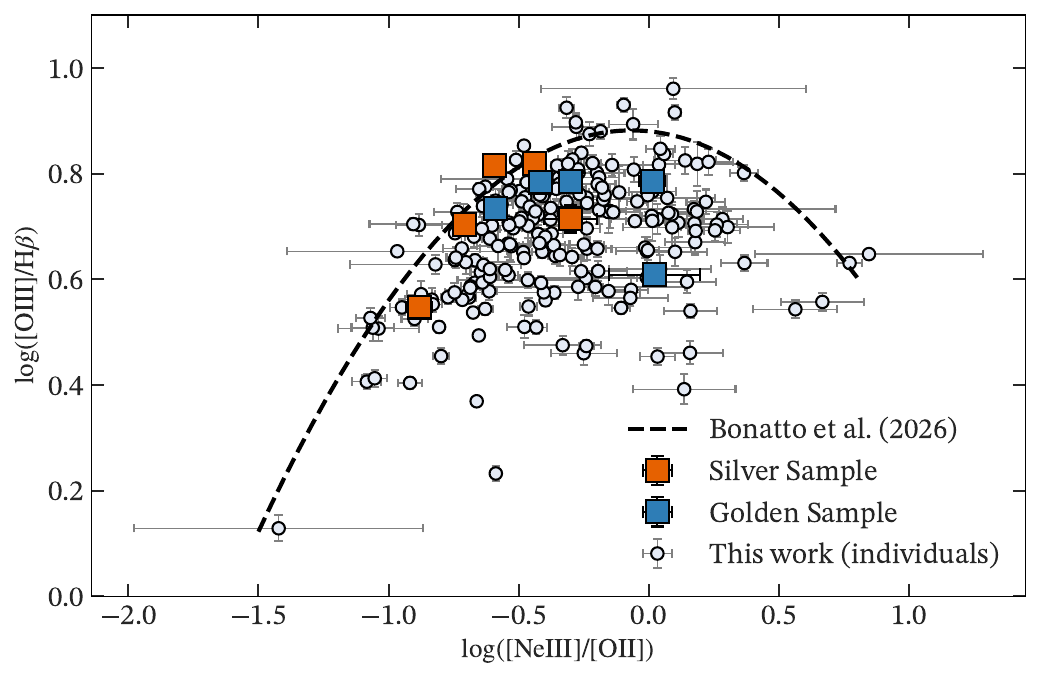}
        \label{fig:Ne3O2_BPT}
    \end{subfigure}
    \caption{\textit{Left}: BPT diagram showing $\log(\mathrm{[O\,III]}/\mathrm{H}\beta)$ 
        versus $\log(\mathrm{[N\,II]}/\mathrm{H}\alpha)$ for the stacked galaxy 
        samples. The Silver sample is shown with orange squares, while the Golden 
        sample (with detected [O\,III] $\lambda4363$) is shown with blue 
        squares. Error bars include full propagation of uncertainties from 
        emission-line flux measurements into logarithmic space. The dashed and 
        solid curves represent the empirical demarcation from 
        \citet{Kauffmann2003} and the theoretical maximum starburst line from 
        \citet{Kewley2001ApJ...556..121K}, respectively. \textit{Right}: Diagnostic diagram of $\log(\mathrm{[O\,III]}/\mathrm{H}\beta)$ 
        versus $\log(\mathrm{[Ne\,III]}/\mathrm{[O\,II]})$. The stacked Silver 
        sample is shown with orange squares, while the Golden sample is shown 
        with blue squares. Individual galaxies are shown as grey circles. 
        The dashed curve shows the separation from \citet{bonatto2026}.}
    \label{fig:BPT_diagrams}
\end{figure}

\section{Photometry}

\begin{table*}[h]
\centering
\caption{Photometric dataset Mosaic v7 from DJA used in this work.}
\label{tab:mosaic_v7_photometry}

\begin{tabular}{lll}
\hline
Telescope & Instrument & Filters \\
\hline
JWST & NIRCam &
F070W, F090W, F115W, F140M, F150W, F150W2, F182M, F200W, F210M, F250M, \\
 &  &
F277W, F322W2, F356W, F360M, F410M, F430M, F444W, F460M, F480M \\[0.3em]

JWST & MIRI &
F770W, F1000W, F1280W, F1500W, F1800W, F2100W \\[0.3em]

HST & ACS/WFC &
F435W, F555W, F606W, F625W, F775W, F814W \\[0.3em]

HST & WFC3/UVIS &
F275W, F336W, F390W, F606W, F814W \\[0.3em]

HST & WFC3/IR &
F105W, F110W, F125W, F140W, F160W \\
\hline
\end{tabular}
\end{table*}
\newpage
\section{Stacked spectra}
\label{sec.stacked}
For each galaxy, we shift the observed spectrum to the rest frame using the redshift of the galaxies based on the [O\,III] lines. All spectra were then registered onto a common wavelength grid, whose sampling is estimated from the median resolution of the galaxies within the bin. Before stacking, each spectrum is normalised by its H$\beta$ flux to avoid more luminous galaxies weighing more in the final stack. Finally, we then generate the composite spectrum as the median of all normalised spectra, applying 3-$\sigma$ clipping and correcting for extinction using the same method described in sect.\ref{sec:dust}. Emission line measurements from each stacked spectrum were done using LiME  \citep{vital2024}, and metallicity was derived following sec.\ref{O/H method}.

\begin{figure}[h]
    \centering
    \includegraphics[width=1\linewidth]{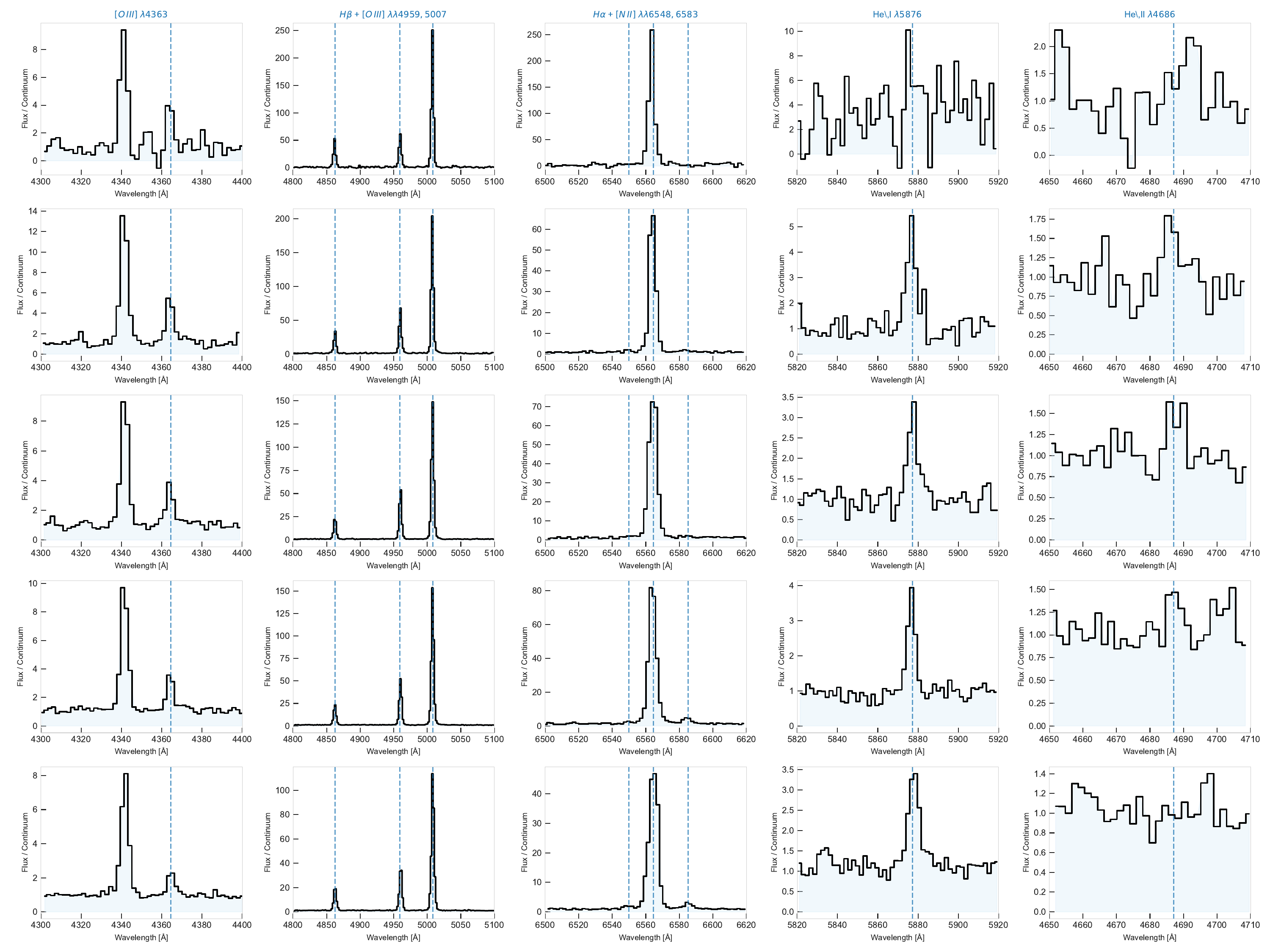}
    \caption{Stacked spectra of galaxies with $\lambda4363$ detections (Golden sample). 
    Rows correspond to increasing stellar mass from top to bottom:
    [7–8], [8–8.5], [8.5–9], [9–9.5], and [9.5–10] $\log(M_\star/M_\odot)$. 
    Vertical dashed lines mark rest-frame emission lines.}
    \label{fig:stack_det}
\end{figure}

\begin{figure}[!t]
    \centering
    \includegraphics[width=1\linewidth]{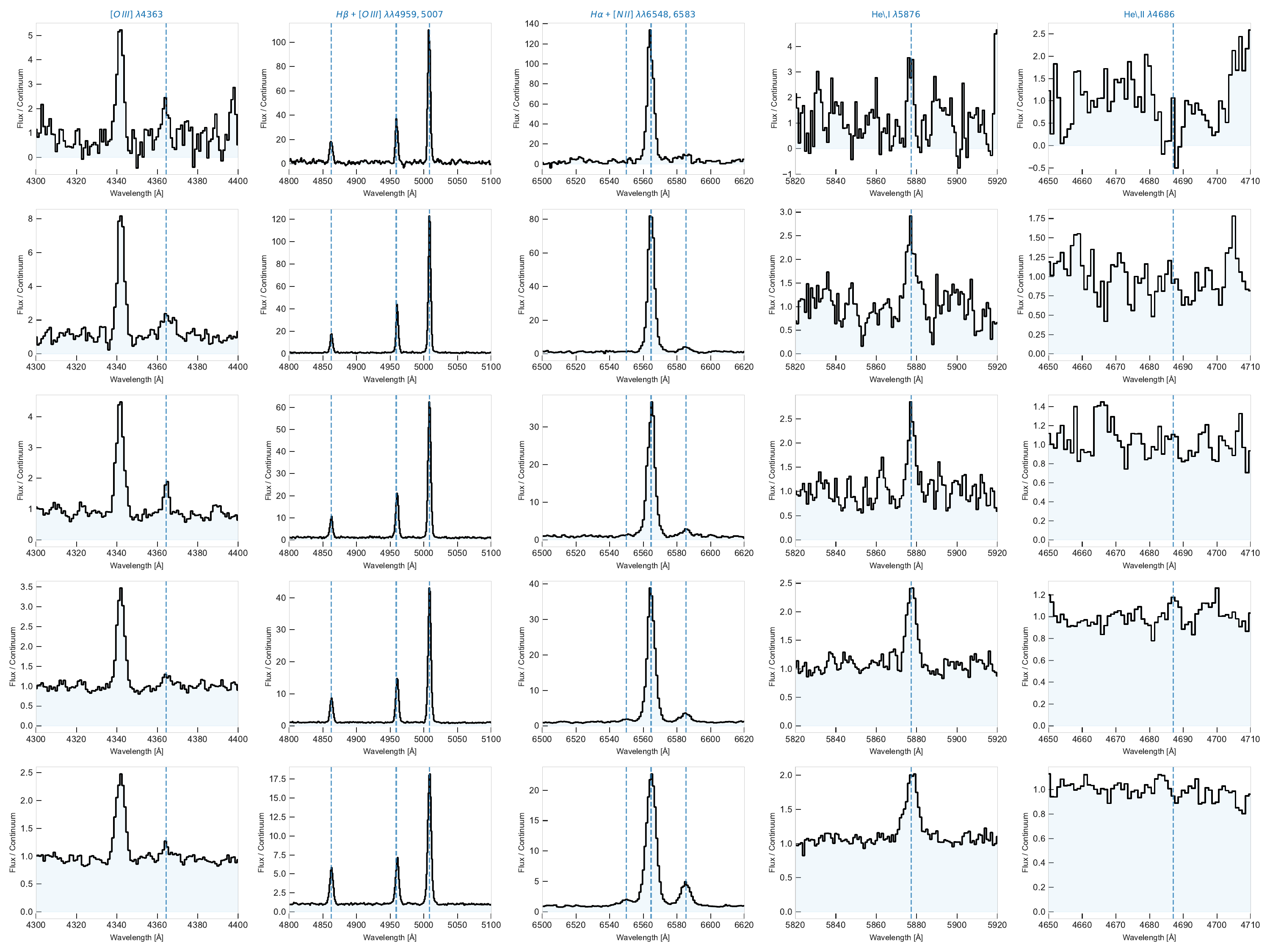}
    \caption{Same as Fig.~\ref{fig:stack_det} but for galaxies without $\lambda4363$ detections (Silver sample).}
    \label{fig:stack_nondet}
\end{figure}

\end{document}